\documentclass{nature_plusfigure}
\usepackage{graphicx}
\usepackage{aas_macros,amsmath,amssymb,xcolor,url}
\usepackage{multirow}
\usepackage{longtable}

\usepackage[nolist]{acronym}

\graphicspath{{./}{figures/}}
\usepackage{lineno}

\usepackage[symbol]{footmisc}

\usepackage[none]{hyphenat}
\usepackage{multibib}
\newcites{New}{References}
\usepackage{cleveref}
\usepackage{lineno}

\title{Discovery and confirmation of the shortest gamma ray burst from a collapsar}

\author{Tom{\'a}s Ahumada$^{1,2,3}$,
	    Leo P. Singer$^{2,4}$,
	    Shreya Anand$^{5}$,
	    Michael W. Coughlin$^{6}$,
	    Mansi M. Kasliwal$^{5}$,
	    Geoffrey Ryan$^{1,2}$,
	    Igor Andreoni$^{5}$,
	    S. Bradley Cenko$^{3,4}$,
	    Christoffer Fremling$^{5}$,
	    Harsh Kumar$^{7,8}$,
	    Peter T. H. Pang$^{9,10}$,
	    Eric Burns$^{11}$,
	    Virginia Cunningham$^{1,2}$,
	    Simone Dichiara$^{1,2}$,
	    Tim Dietrich$^{12}$,
	    Dmitry S. Svinkin$^{13}$,
	    Mouza Almualla$^{14}$,
	    Alberto J. Castro-Tirado$^{15,16}$,
	    Kishalay De$^{5}$,
	    Rachel Dunwoody$^{17}$,
	    Pradip Gatkine$^{5}$,
	    Erica Hammerstein$^{1}$,
	    Shabnam Iyyani$^{18}$,
	    Joseph Mangan$^{17}$,
	    Dan Perley$^{19}$,
	    Sonalika Purkayastha$^{20}$,
	    Eric Bellm$^{21}$, 
        Varun Bhalerao$^{7}$,
        Bryce Bolin$^{5}$,
        Mattia Bulla$^{22}$,
        Christopher Cannella$^{23}$,
        Poonam Chandra$^{20,24}$,
        Dmitry A. Duev$^{5}$,
        Dmitry Frederiks$^{13}$,
        Avishay Gal-Yam$^{25}$,
        Matthew Graham$^{5}$,
        Anna Y. Q. Ho$^{26,27}$,
        Kevin Hurley$^{28}$,
        Viraj Karambelkar$^{5}$,
        Erik C. Kool$^{29}$,
        S. R. Kulkarni$^{5}$,
        Ashish Mahabal$^{5}$,
        Frank Masci$^{30}$,
        Sheila McBreen$^{17}$,
        Shashi B. Pandey$^{31}$,
	    Simeon Reusch$^{32,33}$,
	    Anna Ridnaia$^{13}$,
	    Philippe Rosnet$^{34}$,
	    Benjamin Rusholme$^{30}$,
	    Ana Sagu{\'e}s Carracedo$^{35}$,
	    Roger Smith$^{36}$, 
        Maayane Soumagnac$^{25,37}$,
	    Robert Stein$^{32,33}$,
	    Eleonora Troja$^{3,1}$,
	    Anastasia Tsvetkova$^{13}$,
	    Richard Walters$^{36}$,
	    and Azamat F. Valeev$^{38}$
	}

\newcommand{\arcsec}{$^{\prime\prime}$}

\begin{document}

\maketitle

\begin{affiliations}
\item{Department of Astronomy, University of Maryland, College Park, MD 20742, USA}
\item{Astrophysics Science Division, NASA Goddard Space Flight Center, MC 661, Greenbelt, MD 20771, USA}
\item{Center for Research and Exploration in Space Science and Technology, NASA Goddard Space Flight Center, Greenbelt, MD 20771, USA}
\item{Joint Space-Science Institute, University of Maryland, College Park, MD 20742, USA}
\item{Division of Physics, Mathematics, and Astronomy, California Institute of Technology, Pasadena, CA 91125, USA}
\item{School of Physics and Astronomy, University of Minnesota, Minneapolis, Minnesota 55455, USA}
\item{Indian Institute of Technology Bombay, Powai, Mumbai 400076, India}
\item{LSSTC Data Science Fellow}
\item{Nikhef, Science Park 105, 1098 XG Amsterdam, The Netherlands}
\item{Department of Physics, Utrecht University, Princetonplein 1, 3584 CC Utrecht, The Netherlands}
\item{Louisiana State University, Baton Rouge, LA 70803, USA}
\item{Institut f\"{u}r Physik und Astronomie, Universit\"{a}t Potsdam, Haus 28, Karl-Liebknecht-Str. 24/25, 14476, Potsdam, Germany}
\item{Ioffe Institute, Polytekhnicheskaya, 26, St. Petersburg, 194021 - Russian Federation}
\item{American University of Sharjah, Physics Department, PO Box 26666, Sharjah, UAE}
\item{Instituto de Astrof\'isica de Andaluc\'ia (IAA-CSIC), Glorieta de la Astronom\'ia s/n, E-18008, Granada, Spain}
\item{ Departamento de Ingenier\'ia de Sistemas y Autom\'atica, Escuela de Ingenieros Industriales, Universidad de M\'alaga, Unidad Asociada al CSIC, C. Dr. Ortiz Ramos sn, 29071 M\'alaga, Spain}
\item{School of Physics, University College Dublin, Dublin 4, Ireland}
\item{Inter-University Centre for Astronomy and Astrophysics, Pune, 411007, India}
\item{Astrophysics Research Institute, Liverpool John Moores University,  IC2, Liverpool Science Park, 146 Brownlow Hill, Liverpool L3 5RF, UK}
\item{National Centre for Radio Astrophysics, Tata Institute of Fundamental Research,
Pune University Campus, Ganeshkhind, Pune 411 007, India}
\item{DIRAC Institute, Department of Physics and Astronomy, University of Washington, 3910 15th Avenue NE, Seattle, WA 98195, USA}
\item{Nordita, KTH Royal Institute of Technology and Stockholm University, Roslagstullsbacken 23, 106 91 Stockholm, Sweden}
\item{Duke University, Electrical and Computer Engineering, Durham, NC 27708, USA}
\item{Swarna Jayanti Fellow, Department of Science \& Technology, India}
\item{Department of Particle Physics and Astrophysics, Weizmann Institute of Science, Rehovot, 76100, Israel}
\item{Miller Institute for Basic Research in Science, University of California, Berkeley, CA 94720, USA}
\item{Department of Astronomy, University of California -- Berkeley, Berkeley, CA 94720-5800, USA}
\item{Space Sciences Laboratory, University of California -- Berkeley, Berkeley, CA 94720-7450, USA}
\item{The Oskar Klein Centre, Department of Astronomy, Stockholm University, AlbaNova, SE-106 91 Stockholm, Sweden}
\item{IPAC, California Institute of Technology, 1200 E. California Blvd, Pasadena, CA 91125, USA}
\item{Aryabhatta Research Institute of Observational Sciences (ARIES), Nainital, Uttrakhand, India}
\item{Deutsches Elektronen Synchrotron DESY, Platanenallee 6, 15738 Zeuthen, Germany}
\item{Institut f{\"u}r Physik, Humboldt-Universit{\"a}t zu Berlin, D-12489 Berlin, Germany}
\item{Universit\'e Clermont Auvergne, CNRS/IN2P3, Laboratoire de Physique de Clermont, F-63000 Clermont-Ferrand, France}
\item{The Oskar Klein Centre, Department of Physics, Stockholm University, AlbaNova, SE-106 91 Stockholm, Sweden}
\item{Caltech Optical Observatories, California Institute of Technology, Pasadena, CA 91125, USA}
\item{Lawrence Berkeley National Laboratory, 1 Cyclotron Road, Berkeley, CA 94720, USA}
\item{Special Astrophysical Observatory, Russian Academy of Sciences, Nizhnii Arkhyz, 369167 Russia}
\end{affiliations}

\providecommand{\acrolowercase}[1]{\lowercase{#1}}

\begin{acronym}
\acro{2D}[2D]{two\nobreakdashes-dimensional}
\acro{2+1D}[2+1D]{2+1\nobreakdashes--dimensional}
\acro{2MRS}[2MRS]{2MASS Redshift Survey}
\acro{3D}[3D]{three\nobreakdashes-dimensional}
\acro{2MASS}[2MASS]{Two Micron All Sky Survey}
\acro{AdVirgo}[AdVirgo]{Advanced Virgo}
\acro{AMI}[AMI]{Arcminute Microkelvin Imager}
\acro{AGN}[AGN]{active galactic nucleus}
\acroplural{AGN}[AGN\acrolowercase{s}]{active galactic nuclei}
\acro{aLIGO}[aLIGO]{Advanced \acs{LIGO}}
\acro{ASKAP}[ASKAP]{Australian \acl{SKA} Pathfinder}
\acro{ATCA}[ATCA]{Australia Telescope Compact Array}
\acro{ATLAS}[ATLAS]{Asteroid Terrestrial-impact Last Alert System}
\acro{BAT}[BAT]{Burst Alert Telescope\acroextra{ (instrument on \emph{Swift})}}
\acro{BATSE}[BATSE]{Burst and Transient Source Experiment\acroextra{ (instrument on \acs{CGRO})}}
\acro{BAYESTAR}[BAYESTAR]{BAYESian TriAngulation and Rapid localization}
\acro{BBH}[BBH]{binary black hole}
\acro{BHBH}[BHBH]{\acl{BH}\nobreakdashes--\acl{BH}}
\acro{BH}[BH]{black hole}
\acro{BNS}[BNS]{binary neutron star}
\acro{CARMA}[CARMA]{Combined Array for Research in Millimeter\nobreakdashes-wave Astronomy}
\acro{CASA}[CASA]{Common Astronomy Software Applications}
\acro{CBCG}[CBCG]{Compact Binary Coalescence Galaxy}
\acro{CFH12k}[CFH12k]{Canada--France--Hawaii $12\,288 \times 8\,192$ pixel CCD mosaic\acroextra{ (instrument formerly on the Canada--France--Hawaii Telescope, now on the \ac{P48})}}
\acro{CLU}[CLU]{Census of the Local Universe}
\acro{CRTS}[CRTS]{Catalina Real-time Transient Survey}
\acro{CTIO}[CTIO]{Cerro Tololo Inter-American Observatory}
\acro{CBC}[CBC]{compact binary coalescence}
\acro{CCD}[CCD]{charge coupled device}
\acro{CDF}[CDF]{cumulative distribution function}
\acro{CGRO}[CGRO]{Compton Gamma Ray Observatory}
\acro{CMB}[CMB]{cosmic microwave background}
\acro{CRLB}[CRLB]{Cram\'{e}r\nobreakdashes--Rao lower bound}
\acro{CV}[CV]{Cataclysmic Variable}
\acro{cWB}[\acrolowercase{c}WB]{Coherent WaveBurst}
\acro{DASWG}[DASWG]{Data Analysis Software Working Group}
\acro{DBSP}[DBSP]{Double Spectrograph\acroextra{ (instrument on \acs{P200})}}
\acro{DCT}[DCT]{Discovery Channel Telescope}
\acro{DDT}[DDT]{Director's Discretionary Time}
\acro{DECAM}[DECam]{Dark Energy Camera\acroextra{ (instrument on the Blanco 4\nobreakdashes-m telescope at \acs{CTIO})}}
\acro{DES}[DES]{Dark Energy Survey}
\acro{DFT}[DFT]{discrete Fourier transform}
\acro{EM}[EM]{electromagnetic}
\acro{ER8}[ER8]{eighth engineering run}
\acro{FD}[FD]{frequency domain}
\acro{FAR}[FAR]{false alarm rate}
\acro{FFT}[FFT]{fast Fourier transform}
\acro{FIR}[FIR]{finite impulse response}
\acro{FITS}[FITS]{Flexible Image Transport System}
\acro{F2}[F2]{FLAMINGOS\nobreakdashes-2}
\acro{FLOPS}[FLOPS]{floating point operations per second}
\acro{FOV}[FOV]{field of view}
\acroplural{FOV}[FOV\acrolowercase{s}]{fields of view}
\acro{FTN}[FTN]{Faulkes Telescope North}
\acro{FWHM}[FWHM]{full width at half-maximum}
\acro{GBM}[GBM]{Gamma-ray Burst Monitor\acroextra{ (instrument on \emph{Fermi})}}
\acro{GCN}[GCN]{Gamma-ray Coordinates Network}
\acro{GIT}[GIT]{GROWTH India telescope }
\acro{GLADE}[GLADE]{Galaxy List for the Advanced Detector Era}
\acro{GMOS}[GMOS]{Gemini Multi-Object Spectrograph\acroextra{ (instrument on the Gemini telescopes)}}
\acro{GMRT}[GMRT]{Giant Metrewave Radio Telescope}
\acro{GRB}[GRB]{gamma-ray burst}
\acro{GROWTH}[GROWTH]{Global Relay of Observatories Watching Transients Happen}
\acro{GSC}[GSC]{Gas Slit Camera}
\acro{GSL}[GSL]{GNU Scientific Library}
\acro{GTC}[GTC]{Gran Telescopio Canarias}
\acro{GW}[GW]{gravitational wave}
\acro{GWGC}[GWGC]{Gravitational Wave Galaxy Catalogue}
\acro{HAWC}[HAWC]{High\nobreakdashes-Altitude Water \v{C}erenkov Gamma\nobreakdashes-Ray Observatory}
\acro{HCT}[HCT]{Himalayan Chandra Telescope}
\acro{HEALPix}[HEALP\acrolowercase{ix}]{Hierarchical Equal Area isoLatitude Pixelization}
\acro{HEASARC}[HEASARC]{High Energy Astrophysics Science Archive Research Center}
\acro{HETE}[HETE]{High Energy Transient Explorer}
\acro{HFOSC}[HFOSC]{Himalaya Faint Object Spectrograph and Camera\acroextra{ (instrument on \acs{HCT})}}
\acro{HMXB}[HMXB]{high\nobreakdashes-mass X\nobreakdashes-ray binary}
\acroplural{HMXB}[HMXB\acrolowercase{s}]{high\nobreakdashes-mass X\nobreakdashes-ray binaries}
\acro{HSC}[HSC]{Hyper Suprime\nobreakdashes-Cam\acroextra{ (instrument on the 8.2\nobreakdashes-m Subaru telescope)}}
\acro{IACT}[IACT]{imaging atmospheric \v{C}erenkov telescope}
\acro{IIR}[IIR]{infinite impulse response}
\acro{IMACS}[IMACS]{Inamori-Magellan Areal Camera \& Spectrograph\acroextra{ (instrument on the Magellan Baade telescope)}}
\acro{IMF}[IMF]{initial mass function}
\acro{IMR}[IMR]{inspiral-merger-ringdown}
\acro{IPAC}[IPAC]{Infrared Processing and Analysis Center}
\acro{IPN}[IPN]{InterPlanetary Network}
\acro{IPTF}[\acrolowercase{i}PTF]{intermediate \acl{PTF}}
\acro{IRAC}[IRAC]{Infrared Array Camera}
\acro{ISM}[ISM]{interstellar medium}
\acro{ISS}[ISS]{International Space Station}
\acro{KAGRA}[KAGRA]{KAmioka GRAvitational\nobreakdashes-wave observatory}
\acro{KDE}[KDE]{kernel density estimator}
\acro{KN}[KN]{kilonova}
\acroplural{KN}[KNe]{kilonovae}
\acro{KPED}[KPED]{Kitt Peak Electron multiplying CCD Demonstrator}
\acro{KW}[KW]{Konus-Wind}
\acro{LAT}[LAT]{Large Area Telescope}
\acro{LCO}[LCO]{Las Cumbres Observatory }
\acro{LCOGT}[LCOGT]{Las Cumbres Observatory Global Telescope}
\acro{LDT}[LDT]{Lowell Discovery Telescope}
\acro{LGRB}[LGRB]{long \acl{GRB}}
\acro{LHO}[LHO]{\ac{LIGO} Hanford Observatory}
\acro{LIB}[LIB]{LALInference Burst}
\acro{LIGO}[LIGO]{Laser Interferometer \acs{GW} Observatory}
\acro{llGRB}[\acrolowercase{ll}GRB]{low\nobreakdashes-luminosity \ac{GRB}}
\acro{LLOID}[LLOID]{Low Latency Online Inspiral Detection}
\acro{LLO}[LLO]{\ac{LIGO} Livingston Observatory}
\acro{LMI}[LMI]{Large Monolithic Imager\acroextra{ (instrument on \ac{DCT})}}
\acro{LOFAR}[LOFAR]{Low Frequency Array}
\acro{LOS}[LOS]{line of sight}
\acroplural{LOS}[LOSs]{lines of sight}
\acro{LMC}[LMC]{Large Magellanic Cloud}
\acro{LRIS}[LRIS]{Low Resolution Imaging Spectrograph}
\acro{LS}[LS]{Legacy Survey}
\acro{LSB}[LSB]{long, soft burst}
\acro{LSC}[LSC]{\acs{LIGO} Scientific Collaboration}
\acro{LSO}[LSO]{last stable orbit}
\acro{LSST}[LSST]{Large Synoptic Survey Telescope}
\acro{LT}[LT]{Liverpool Telescope}
\acro{LTI}[LTI]{linear time invariant}
\acro{MAP}[MAP]{maximum a posteriori}
\acro{MBTA}[MBTA]{Multi-Band Template Analysis}
\acro{MCMC}[MCMC]{Markov chain Monte Carlo}
\acro{Methods}[Methods]{Methods}
\acro{MLE}[MLE]{\ac{ML} estimator}
\acro{ML}[ML]{maximum likelihood}
\acro{MPC}[MPC]{Minor Planet Center}
\acro{MOU}[MOU]{memorandum of understanding}
\acroplural{MOU}[MOUs]{memoranda of understanding}
\acro{MWA}[MWA]{Murchison Widefield Array}
\acro{NED}[NED]{NASA/IPAC Extragalactic Database}
\acro{NIR}[NIR]{near infrared}
\acro{NSBH}[NSBH]{neutron star\nobreakdashes--black hole}
\acro{NSBH}[NSBH]{\acl{NS}\nobreakdashes--\acl{BH}}
\acro{NSF}[NSF]{National Science Foundation}
\acro{NSNS}[NSNS]{\acl{NS}\nobreakdashes--\acl{NS}}
\acro{NS}[NS]{neutron star}
\acro{O1}[O1]{\acl{aLIGO}'s first observing run}
\acro{O2}[O2]{\acl{aLIGO}'s second observing run}
\acro{O3}[O3]{\acl{aLIGO}'s and \acl{AdVirgo} third observing run}
\acro{oLIB}[\acrolowercase{o}LIB]{Omicron+\acl{LIB}}
\acro{OT}[OT]{optical transient}
\acro{P48}[P48]{Palomar 48~inch Oschin telescope}
\acro{P60}[P60]{robotic Palomar 60~inch telescope}
\acro{P200}[P200]{Palomar 200~inch Hale telescope}
\acro{PC}[PC]{photon counting}
\acro{PESSTO}[PESSTO]{Public ESO Spectroscopic Survey of Transient Objects}
\acro{pPXF}[pPXF]{Penalized Pixel-Fitting}
\acro{PSD}[PSD]{power spectral density}
\acro{PSF}[PSF]{point-spread function}
\acro{PS1}[PS1]{Pan\nobreakdashes-STARRS~1}
\acro{PTF}[PTF]{Palomar Transient Factory}
\acro{QUEST}[QUEST]{Quasar Equatorial Survey Team}
\acro{RAPTOR}[RAPTOR]{Rapid Telescopes for Optical Response}
\acro{REU}[REU]{Research Experiences for Undergraduates}
\acro{RMS}[RMS]{root mean square}
\acro{ROTSE}[ROTSE]{Robotic Optical Transient Search}
\acro{S5}[S5]{\ac{LIGO}'s fifth science run}
\acro{S6}[S6]{\ac{LIGO}'s sixth science run}
\acro{SAA}[SAA]{South Atlantic Anomaly}
\acro{SHB}[SHB]{short, hard burst}
\acro{SHGRB}[SHGRB]{short, hard \acl{GRB}}
\acro{SI}[SI]{Supplementary Information}
\acro{SKA}[SKA]{Square Kilometer Array}
\acro{SMT}[SMT]{Slewing Mirror Telescope\acroextra{ (instrument on \acs{UFFO} Pathfinder)}}
\acro{SNR}[S/N]{signal\nobreakdashes-to\nobreakdashes-noise ratio}
\acro{SEDM}[SEDM]{Spectral Energy Distribution Machine}
\acro{SSC}[SSC]{synchrotron self\nobreakdashes-Compton}
\acro{SDSS}[SDSS]{Sloan Digital Sky Survey}
\acro{SED}[SED]{spectral energy distribution}
\acro{SFR}[SFR]{star formation rate}
\acro{SGRB}[SGRB]{short \acl{GRB}}
\acro{SN}[SN]{supernova}
\acroplural{SN}[SN\acrolowercase{e}]{supernova}
\acro{SNIa}[\acs{SN}\,I\acrolowercase{a}]{Type~Ia \ac{SN}}
\acroplural{SNIa}[\acsp{SN}\,I\acrolowercase{a}]{Type~Ic \acp{SN}}
\acro{SNIcBL}[\acs{SN}\,I\acrolowercase{c}\nobreakdashes-BL]{broad\nobreakdashes-line Type~Ic \ac{SN}}
\acroplural{SNIcBL}[\acsp{SN}\,I\acrolowercase{c}\nobreakdashes-BL]{broad\nobreakdashes-line Type~Ic \acp{SN}}
\acro{SVD}[SVD]{singular value decomposition}
\acro{TAROT}[TAROT]{T\'{e}lescopes \`{a} Action Rapide pour les Objets Transitoires}
\acro{TDOA}[TDOA]{time delay on arrival}
\acroplural{TDOA}[TDOA\acrolowercase{s}]{time delays on arrival}
\acro{TD}[TD]{time domain}
\acro{TOA}[TOA]{time of arrival}
\acroplural{TOA}[TOA\acrolowercase{s}]{times of arrival}
\acro{ToO}[T\acrolowercase{o}O]{Target\nobreakdashes-of\nobreakdashes-Opportunity}
\acroplural{TOO}[TOO\acrolowercase{s}]{targets of opportunity}
\acro{TNS}[TNS]{Transit Name Server}
\acro{UFFO}[UFFO]{Ultra Fast Flash Observatory}
\acro{UHE}[UHE]{ultra high energy}
\acro{UVOT}[UVOT]{UV/Optical Telescope\acroextra{ (instrument on \emph{Swift})}}
\acro{VHE}[VHE]{very high energy}
\acro{VISTA}[VISTA@ESO]{Visible and Infrared Survey Telescope}
\acro{VLA}[VLA]{Karl G. Jansky Very Large Array}
\acro{VLT}[VLT]{Very Large Telescope}
\acro{VST}[VST@ESO]{\acs{VLT} Survey Telescope}
\acro{WAM}[WAM]{Wide\nobreakdashes-band All\nobreakdashes-sky Monitor\acroextra{ (instrument on \emph{Suzaku})}}
\acro{WCS}[WCS]{World Coordinate System}
\acro{WISE}[WISE]{Wide-field Infrared Survey Explorer}
\acro{WSS}[w.s.s.]{wide\nobreakdashes-sense stationary}
\acro{XRF}[XRF]{X\nobreakdashes-ray flash}
\acroplural{XRF}[XRF\acrolowercase{s}]{X\nobreakdashes-ray flashes}
\acro{XRT}[XRT]{X\nobreakdashes-ray Telescope\acroextra{ (instrument on \emph{Swift})}}
\acro{ZTF}[ZTF]{Zwicky Transient Facility}
\end{acronym}

\begin{abstract}

Gamma-ray bursts (\acsp{GRB}\acused{GRB}) are among the brightest and most energetic events in the universe. The duration and hardness distribution of GRBs has two clusters\cite{KoMe93}, now understood to reflect (at least) two different progenitors\cite{nakar07}. Short-hard GRBs (\acsp{SGRB}\acused{SGRB}; $T_{90}<$2\,s) arise from compact binary mergers, while long-soft GRBs (\acsp{LGRB}\acused{LGRB}; $T_{90}>$2\,s) have been attributed to the collapse of peculiar massive stars (collapsars)\cite{Woosley2006}.  The discovery of SN\,1998bw/GRB\,980425\cite{galama1998unusual} marked the first association of a \ac{LGRB} with a collapsar and AT\,2017gfo\cite{2017Sci...358.1556C}/GRB\,170817A/GW170817\cite{gold17} marked the first association of a \ac{SGRB} with a binary neutron star merger, producing also \ac{GW}. 
Here, we present the discovery of ZTF20abwysqy (AT2020scz), a fast-fading optical transient in the \emph{Fermi} Satellite and the \ac{IPN} localization regions of GRB\,200826A; X-ray and radio emission further confirm that this is the afterglow. 
Follow-up imaging (at rest-frame 16.5 days) reveals excess emission above the afterglow that cannot be explained as an underlying \ac{KN}, but is consistent with being the \ac{SN}.
Despite the GRB duration being short (rest-frame $T_{90}$ of 0.65\,s), our panchromatic follow-up data confirms a collapsar origin. 
GRB\,200826A is the shortest \ac{LGRB} found with an associated collapsar; it appears to sit on the brink between a successful and a failed collapsar. Our discovery is consistent with 
the hypothesis that most collapsars fail to produce ultra-relativistic jets. 
\end{abstract}


On August 26, 2020, at 04:29:52 UT, the \ac{GBM} on board the \textit{Fermi} Gamma-ray Space Telescope detected GRB\,200826A with duration ($T_{90}$) of $1.14\pm 0.13$ seconds in the 50--300 keV energy range
In addition to \emph{Fermi} (\ac{GBM} trigger 620108997), GRB\,200826A was detected by four other Interplanetary Network (IPN) instruments (see Methods\acused{Methods}).

\begin{figure*}
    \centering
    \includegraphics[width=\textwidth]{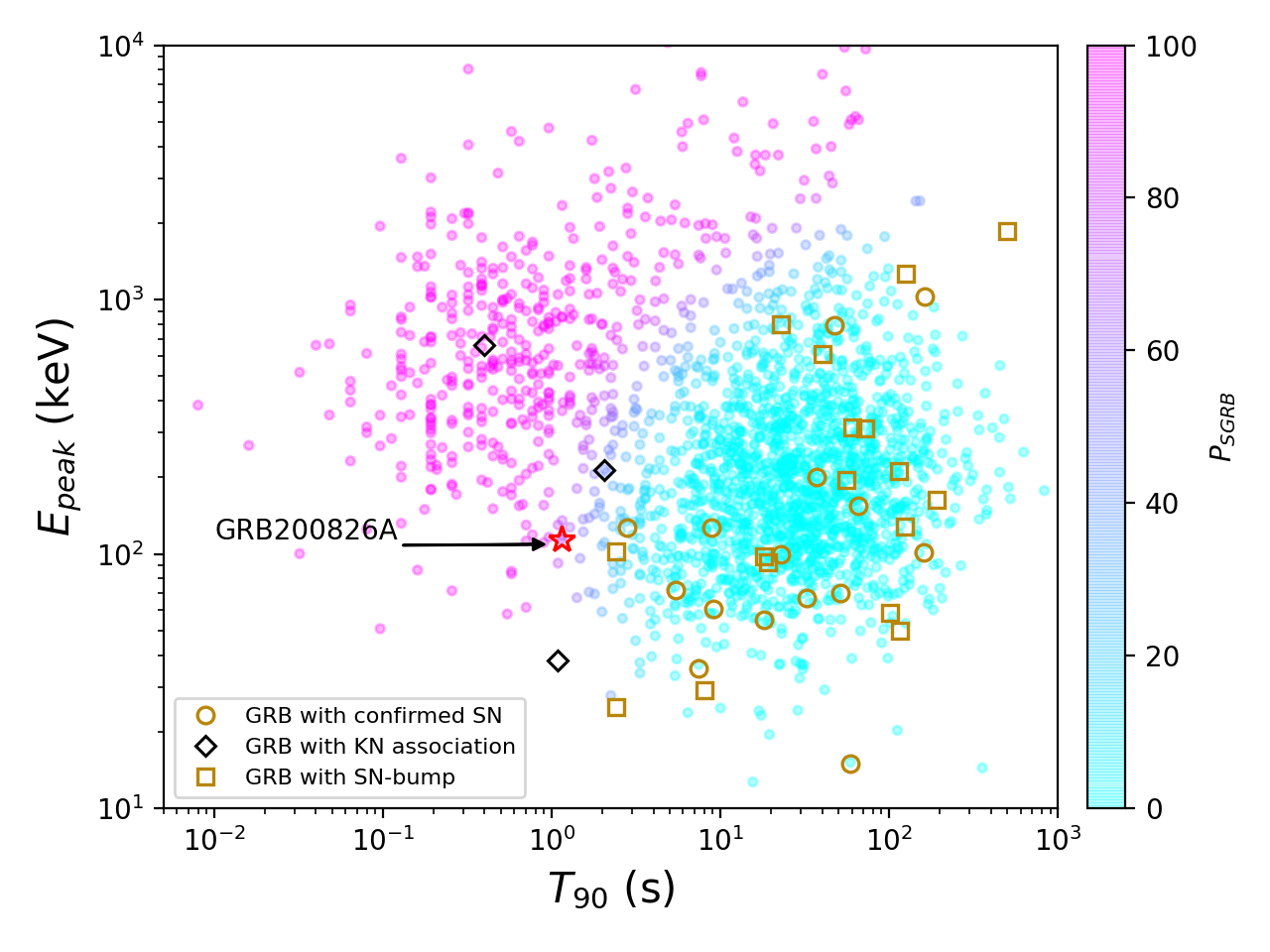}
\caption{\label{fig:gamma-rays} \textbf{Gamma-ray properties of GRB\,200826A in context.}    
The peak energy based on a Comptonised fit, $E_{peak}$ (keV), vs. the time-integrated, T$_{90}$ (s), for 2310 \textit{Fermi} GBM GRBs. The data are fit with two log-normal distributions for the two GRB classes. The color of the data points indicate the probability with magenta being 100\% SGRB and cyan being 100\% LGRB. GRB\,200826A is surrounded by a red star with a SGRB probability of 74\% (See Methods). Yellow squares show LGRBs with SN-bumps, yellow circles show LGRBs with spectroscopically confirmed SN\cite{cano2017}, black diamonds show SGRBs with claimed \ac{KN} excess\cite{Gompertz2018}.
}
\end{figure*}

The gamma-ray properties alone do not always yield an unambiguous classification. Some \acp{SGRB} show afterglow and host properties akin to \acp{LGRB}, e.g. Ref.~\cite{antonelli2009grb}; 
some \acp{LGRB} show no evidence for collapsars to deep limits akin to \acp{SGRB}, e.g.\ Ref.~\cite{GalYam2006}. 
Based solely on $T_{90}$\cite{BrNa2013}, GRB\,200826A has a \ac{SGRB} probability of $65\%^{+12}_{-11}$. Also taking into consideration the $E_{peak}$ parameter of a Comptonized model fit to the single spectrum over the duration of the burst (see \ref{tab:timeres_gbm}), the probability that GRB\,200826A is a \ac{SGRB} increases to 74\% (see Figure~\ref{fig:gamma-rays} and Methods). However, based on rest-frame energetics, GRB\,200826A is not consistent with the \ac{SGRB} population (right panel in~\ref{fig:grb_lc}). 

\begin{figure*}
\includegraphics[width=\textwidth]{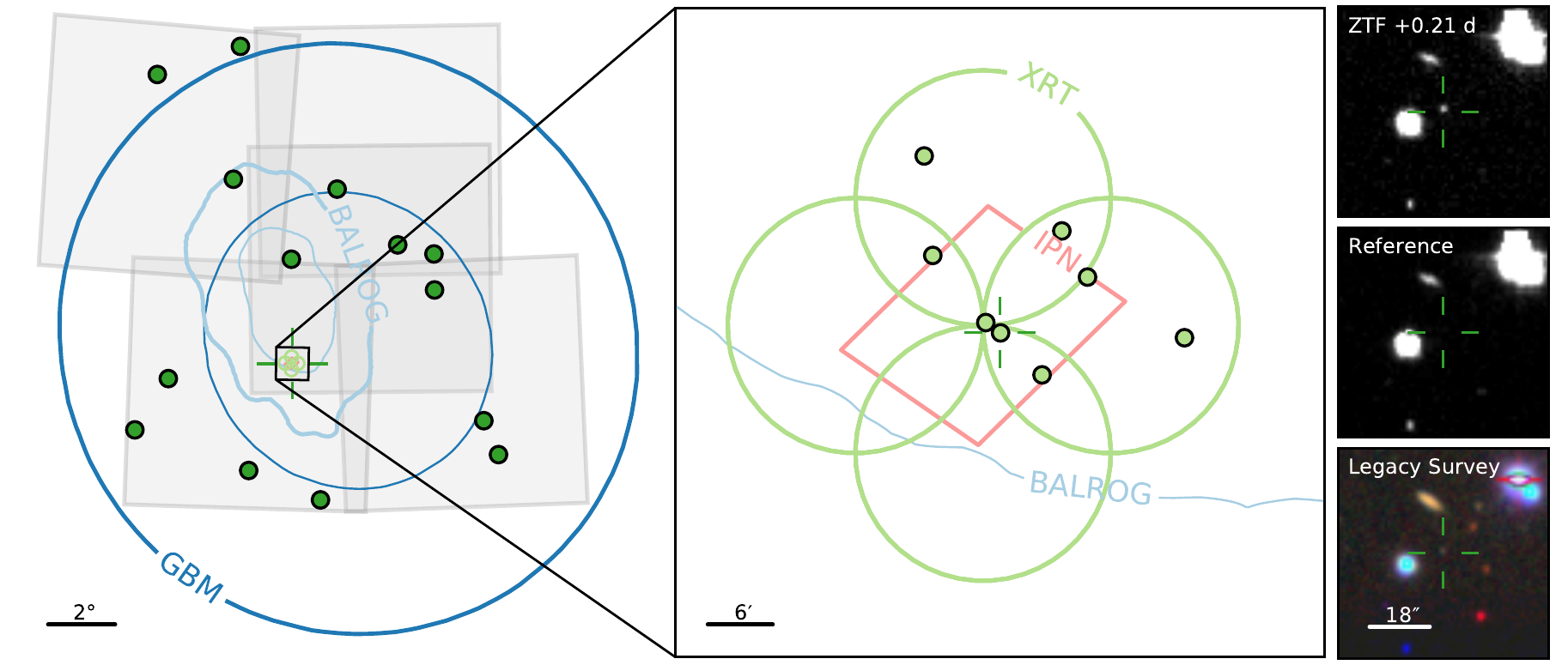}
\caption{\label{fig:1Discovery}\textbf{Discovery of the afterglow of GRB\,200826A.} It was found at the position $\alpha= 00^h27^m08.542^s$, $\delta =+34^h01^m38.327^s$ (J2000) with an uncertainty of 0.08.'' Contours in the left panel represent the \emph{Fermi} \ac{GBM} 90\% (thick) and 50\% (thin) credible regions from the official \emph{Fermi} \ac{GBM} localization (dark blue) and BALROG (light blue). The filled gray squares show the ZTF fields observed and the dark green dots are the positions of the ZTF Night 1 optical candidates. In the middle panel, $3\sigma$ IPN triangulation is shown in pink; the four fields of the four XRT tiled observations are shown as light green circles and the XRT candidates are light green dots. The position of the afterglow is marked by the dark green reticle. The right-hand panels are centered at the position of ZTF20abwysqy. The cutouts, from top to bottom, show the ZTF discovery image, the ZTF stacked reference image, and a false color image showing the host galaxy from Legacy Survey DR8. In the cutouts, North is up and East to the left.}
\end{figure*}

Starting 4.2\,hours after the \ac{GRB}, we observed 180 sq. degrees of the \emph{Fermi}-\ac{GBM} localization with the \ac{ZTF}\cite{bellm2018zwicky} (see Fig.~\ref{fig:1Discovery}). 
At 17\,hours after the \ac{GRB}, IPN triangulated the source to a smaller region.
ZTF20abwysqy was the only candidate that passed our alert filtering scheme and was also inside the \ac{IPN} region (see Methods). We discovered ZTF20abwysqy at a brightness of $g=20.86\pm0.04$ mag (AB system). The previous upper limit ($g > 21.3$\,mag at $5\sigma$) was 17.3\,hours before the \ac{GRB}, as part of the nominal all-sky survey mode. 
In addition to the spatial coincidence, ZTF20abwysqy was associated with a fading X-ray counterpart\cite{GCN28300} and variable radio emission\cite{GCN28302}, confirming ZTF20abwysqy as the afterglow of GRB\,200826A (see \ref{table:observations_afterglow},\ref{table:observations_xrt} and \ref{table:observations_radio}).
 






ZTF20abwysqy was discovered in a galaxy with archival detections in \ac{PS1}\cite{2016arXiv161205560C} and \ac{LS}\cite{Dey2019}, at a \ac{LS} photometric redshift of $z_p=0.71 \pm 0.14$. The offset between the host galaxy's centroid and the transient is $0.18\pm0.05$\arcsec, corresponding to a physical distance of $2.09\pm0.6$\,kpc. We acquired a \ac{GTC} spectrum of the galaxy; we see strong [OII] and [OIII] lines at $z=0.748$ (see \ref{table:observations_host} and \ref{fig:galaxySED}). 
With both spectral and photometric \ac{SED} fitting, we infer a stellar mass for the host galaxy of $ \sim10^{9.7}$ \(M_\odot\) (see \ref{fig:galaxySED}, \ref{fig:corner_sed} and \ref{tab:fluxes} in \ac{Methods}), 
which is near the distribution peak for \acp{LGRB}\cite{Leibler2010}, while below the median for \acp{SGRB} based on Ref.\cite{Fong15}.

 We model the GRB afterglow using the standard synchrotron fireball model to constrain parameters related to the energy and geometry of the GRB central engine (see \ac{Methods}). Electrons in the circumburst medium are accelerated by the shock wave and reach a power-law energy distribution characterized by the index $p$, $N(E) \propto E^{-p}$. This results in a \ac{SED} described by a series of broken power laws, e.g. Ref.~\cite{sari98} (See~\ref{fig:SED_afterglow} and \ref{fig:corner_eps}). 
 The associated isotropic kinetic energy of $E_{\text{K,iso}} = 6.0^{+51.3}_{-4.4} \times 10^{52}$ erg lies in the top 5\% of the $E_{K,iso}$ distribution for \acp{SGRB}\cite{Fong15}, but within the 90\% confidence range of the \ac{LGRB} energy distribution\cite{shivvers2011beaming} (see \ac{Methods}). Our data can only loosely constrain the circumburst density. The upper end of the distribution is consistent with the values found for the median circumburst densities\cite{panaitescu2002properties} of \acp{LGRB}, while the lower end is more representative of \acp{SGRB}\cite{Fong15}.

\begin{figure*}[!htb]
    \includegraphics[width=\textwidth]{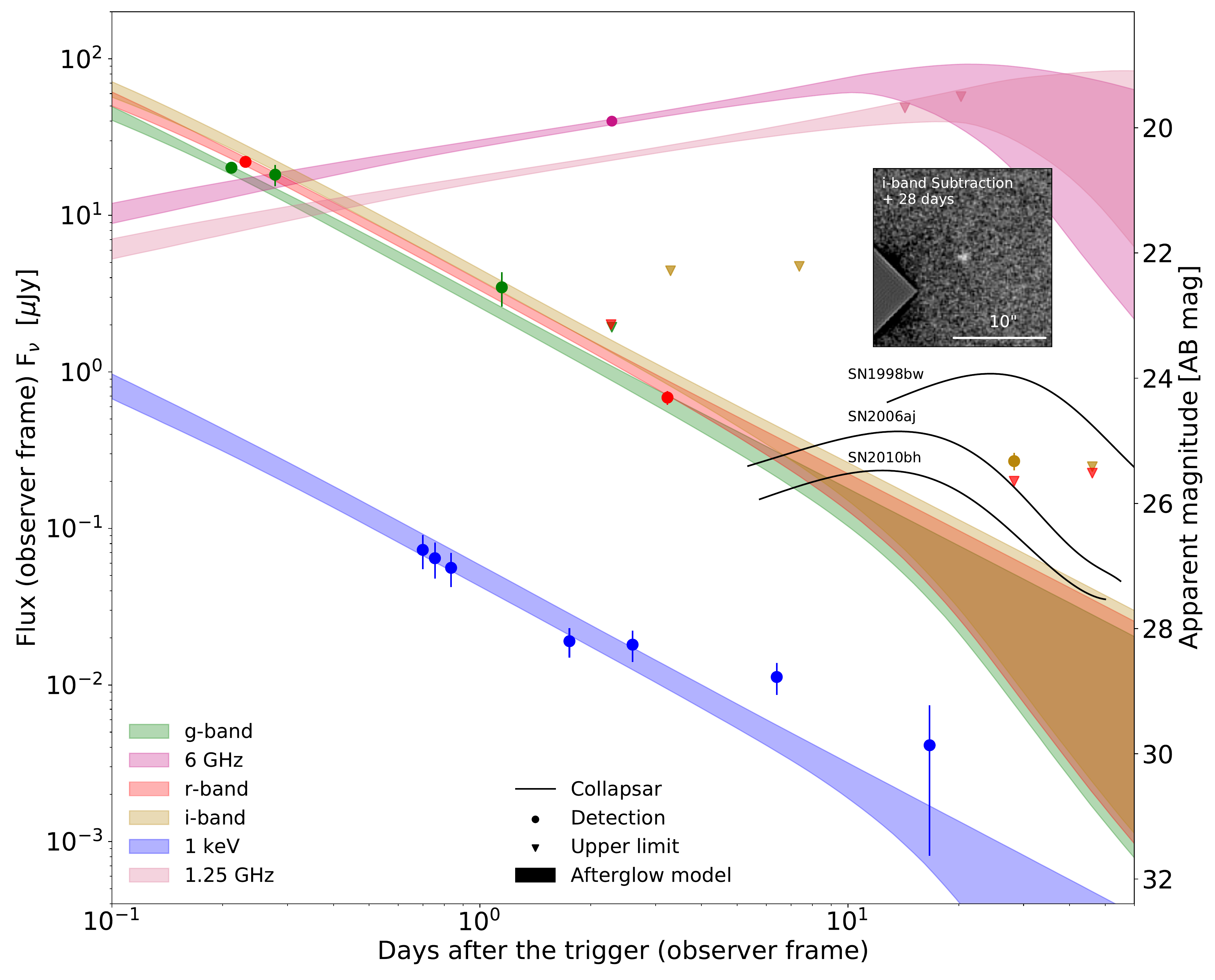}
\caption{\label{fig:lightcurve}\textbf{Panchromatic afterglow and collapsar confirmation.}   
The available multiwavelength light curve data is over-plotted with the best fits from the \texttt{afterglowpy} modeling assuming an ISM-like environment. Detections are shown as circles with their respective error bars, and upper limits are shown as inverted triangles. The optical $g$-, $r$- and $i$- bands are shown in green, red and yellow, the XRT 1\,keV data is shown in blue, the VLA in fuchsia, while the GMRT data is presented in pink. We show the K-corrected light curves of three well-studied GRB-\acp{SN} (SN1998bw, one of the brightest; SN2006aj and SN2010bh, two of the faintest) with solid black lines. The Gemini \ac{GMOS}-N $i$-band detection is shown at day 28.28, as a yellow circle, and is consistent with the collapsar population. We show a cutout of the $i$-band ZOGY subtraction, revealing our 25.45\,mag detection with a ZOGY corrected score of~4.
}
\end{figure*}

In the collapsar scenario, a high-velocity stripped-envelope \ac{SN} (\ac{SN} Ic-BL) should follow the \ac{GRB} detection\cite{Woosley2006}. 
To test this scenario, 
we used the Gemini Multi-Object Spectrograph (GMOS-N\acused{GMOS}; see \ac{Methods}) to acquire $r$- and $i$-band images of ZTF20abwysqy on three different epochs: $\sim$28, $\sim$45, and $\sim$80 days after the \ac{GRB} trigger (epoch 1, 2 and 3 respectively). Using epoch 3 as the reference, we undertook image subtraction using two different subtraction algorithms (see \ac{Methods} for details).
On epoch 1, our Gemini observations show evidence of a transient with an $i$-band magnitude of 25.45$\pm$0.15\,mag (with an $i$-band 5$\sigma$ limit of 25.9\,mag, see \ref{table:observations_afterglow}); there is no source in the $r$-band observations up to a 5$\sigma$ limit of 25.4\,mag (see Fig. \ref{fig:lightcurve}). On epoch 2, we do not detect a source up to a 5$\sigma$ limit of 25.5\,mag in the $i$-band and 25.7 mag in the $r$-band. Thus, at a rest-frame time of $\sim$16 days, the foreground extinction-corrected absolute magnitude of ZTF20abwysqy is $M_i = -18.0$\,mag. 

To understand the source of the $i$-band excess, we use Markov chain Monte Carlo (MCMC) and full forward modeling of all multi-band observations excluding the GMOS detection to compare three scenarios: an afterglow only, an afterglow plus a \ac{KN}, and an afterglow plus a \ac{SN}. The \ac{KN} model is based on a best-fit template to AT2017gfo\cite{DiCo2020} scaled by a compilation of SGRB-\ac{KN} candidates\cite{Gompertz2018}; the collapsar model uses a SN1998bw template\cite{clocchiatti2011ultimate} with stretch and scale parameters drawn from a prior that is consistent with the historical GRB-SN sample\cite{cano2017}. We dismiss the afterglow-only and afterglow-plus-\ac{KN} models because they predict $i$-band flux at the time of the GMOS data point that is too faint by ${1.6}^{+1.8}_{-0.3}$\,mag -- inconsistent with the observations at the $\sim$5$\sigma$ level (see \ref{fig:posterior_prediction_afterglow_supernova} and \ref{tab:pars}).

We repeated the analysis of all three scenarios including the GMOS data point in order to do Bayesian model comparison. The Bayes factor between the afterglow-plus-\ac{KN} and afterglow-only models is $\sim 1$, indicating that neither model is favored over the other. However, the Bayes factor between the afterglow-plus-\ac{SN} and afterglow-only model is $10^{5.5}$, indicating that the afterglow-plus-\ac{SN} model is strongly favored.
Based on the compilation of Ref.~\cite{cano2017} of GRBs with associated \acp{SN}, GRB 200826A is the shortest \ac{LGRB} found with an associated collapsar, with an observed $T_{90}$ of 1.13\,s and rest-frame $T_{90}$ of 0.65\,s (see Figure~\ref{fig:gamma-rays}). The second shortest LGRB with a \ac{SN} is GRB 040924, with a rest-frame $T_{90}$ of 1.29\,s\cite{cano2017}. 

In the conventional fireball model of a \ac{LGRB}, the rest-frame duration of the prompt emission, $t_\gamma$, is the difference between the duration of the activity of the central engine, $t_e$, and the time required to break out of the envelope, $t_b$\cite{sobacchi2017common}. The short duration of GRB 200826A suggests that $t_e \gtrsim t_b$, which might imply that its central engine is active only very briefly, or that the stellar envelope is unusually thick, as compared to other \acp{LGRB}, perhaps requiring a nonstandard progenitor\cite{zhanginprep}. Then it is natural to infer that there must be collapsars for which $t_e < t_b$ that fail to produce a fireball. For $t_e \lesssim t_b$, the jet may fail to clear a path for itself and remains cocooned within the exploding star, yet the cocoon itself may still produce a mildly relativistic shock breakout that manifests as a soft, quasi-thermal, \ac{llGRB} like GRB\,980425/SN\,1998bw or GRB\,060218/SN2006aj\cite{2006Natur.442.1014S,2015ApJ...807..172N}. For $t_e \ll t_b$, there may be no prompt emission at all.

GRB\,200826A, then, may sit on the brink between a successful \ac{LGRB} and a failed one. The sign of one continuous physical parameter, $t_e - t_b$, switches a collapsar-powered GRB discontinuously between two dominant emission mechanisms: internal shocks in a relativistic jet or mildly relativistic shock breakout at the surface of the exploding star. These two mechanisms correspond to two widely separated and disconnected regions in the hardness-duration-fluence phase space: the traditional border between \acp{SGRB} and \acp{LGRB}, and the exceptionally long and soft region occupied by \acp{llGRB} and \acp{XRF}. The local rate of \acp{LGRB} ($\sim$1\,Gpc$^{-3}$yr$^{-1}$)\cite{2015ApJ...807..172N} is 2 orders of magnitude lower than the rate of llGRBs (300\,Gpc$^{-3}$yr$^{-1}$)\cite{2015ApJ...807..172N} and 3 orders of magnitude lower than the rate of \acp{SN} Ic-BL (4500\,Gpc$^{-3}$yr$^{-1}$)\cite{2016ApJ...823..154G}. It is understood that \acp{LGRB}, and to a lesser extent, \acp{llGRB}, are collimated and relativistically beamed, suppressing the detection of events in which the jet is pointed away from our line of sight. However, even allowing for a beaming correction to their rates of 100 for \acp{LGRB} and 1--10 for \acp{llGRB} \cite{2015ApJ...807..172N} (but no beaming correction for \acp{SN}), \acp{LGRB} still occur at a rate (100\,Gpc$^{-3}$yr$^{-1}$) that is up to an order of magnitude lower than that of \acp{llGRB} (300-3000\,Gpc$^{-3}$yr$^{-1}$) or that of \acp{SN} Ic-BL, suggesting that majority of the collapsars fail to produce an ultra-relativistic jet and instead drive a wide-angle or nearly isotropic and only mildly-relativistic cocoon.

Based on indirect evidence from the host, afterglow, and gamma-ray properties, it has been argued that as many as 84\% of bursts detected by the \textit{Neil Gehrels Swift Observatory}\cite{swift} (\emph{Swift}) and 40\% of the \emph{Fermi} bursts that are nominally \acp{SGRB} ($T_{90} < 2$\,s) are actually misclassified \acp{LGRB}\cite{2013RSPTA.37120273P}. If this is correct, then one would expect many more short-duration \acp{GRB} with collapsars that are on the edge between success and failure of the jet. The discovery of GRB\,200826A, an \ac{SGRB} imposter
, lends credence to this bold claim, and would suggest that the rates of such short-duration \acp{LGRB} is comparable to the rate of \acp{llGRB}, up to a few hundred Gpc$^{-3}$yr$^{-1}$. Thus, our discovery upholds the hypothesis that most collapsars fail to produce jets.

\newpage


\clearpage
\bibliographystyle{naturemag}
\bibliography{references}

\begin{addendum}

\item This work was supported by the GROWTH (Global Relay of Observatories Watching Transients Happen) project funded by the National Science Foundation under PIRE Grant No 1545949. GROWTH is a collaborative project among California Institute of Technology (USA), University of Maryland College Park (USA), University of Wisconsin Milwaukee (USA), Texas Tech University (USA), San Diego State University (USA), University of Washington (USA), Los Alamos National Laboratory (USA), Tokyo Institute of Technology (Japan), National Central University (Taiwan), Indian Institute of Astrophysics (India), Indian Institute of Technology Bombay (India), Weizmann Institute of Science (Israel), The Oskar Klein Centre at Stockholm University (Sweden), Humboldt University (Germany), Liverpool John Moores University (UK) and University of Sydney (Australia). 

Based on observations obtained with the Samuel Oschin Telescope 48-inch and the 60-inch Telescope at the Palomar Observatory as part of the Zwicky Transient Facility project. ZTF is supported by the National Science Foundation under Grant No. AST-1440341 and a collaboration including Caltech, IPAC, the Weizmann Institute for Science, the Oskar Klein Center at Stockholm University, the University of Maryland, the University of Washington (UW), Deutsches Elektronen-Synchrotron and Humboldt University, Los Alamos National Laboratories, the TANGO Consortium of Taiwan, the University of Wisconsin at Milwaukee, and Lawrence Berkeley National Laboratories. Operations are conducted by Caltech Optical Observatories, IPAC, and UW.  The work is partly based on the observations made with the Gran Telescopio Canarias (GTC), installed in the Spanish Observatorio del Roque de los Muchachos of the Instituto de Astrofisica de Canarias, in the island of La Palma.

The material is based upon work supported by NASA under award number 80GSFC17M0002.

AJCT acknowledges all co-Is of the GTC proposal and the financial support from the State Agency for Research of the Spanish MCIU through the
``Center of Excellence Severo Ochoa'' award to the Instituto de Astrofísica de Andalucía (SEV-2017-0709).
 
The ZTF forced-photometry service was funded under the Heising-Simons Foundation grant \#12540303 (PI: Graham).

S.McB. and J.M. acknowledge support from Science Foundation Ireland under grant number 17/CDA/4723. R.D.  acknowledges  support  from  the  Irish  Research  Council (IRC) under grant GOIPG/2019/2033

Analysis was performed on the YORP cluster administered by the Center for Theory and Computation, part
of the Department of Astronomy at the University of Maryland.

Resources supporting this work were provided by the NASA High-End Computing (HEC) Program through the NASA Advanced Supercomputing (NAS) Division at Ames Research Center.

These results also made use of Lowell Observatory’s Lowell Discovery Telescope (LDT), formerly the Discovery Channel Telescope. Lowell operates the LDT in partnership
with Boston University, Northern Arizona University, the University of Maryland, and the University of Toledo. Partial support of
the LDT was provided by Discovery Communications. LMI was
built by Lowell Observatory using funds from the National Science Foundation (AST-1005313). 

M.~W.~Coughlin acknowledges support from the National Science Foundation with grant number PHY-2010970.
S.~Anand gratefully acknowledges support from the GROWTH PIRE grant (1545949).
Part of this research was carried out at the Jet Propulsion Laboratory, California Institute of Technology, under a contract with the National Aeronautics and Space Administration. 
E.C. Kool acknowledges support from the G.R.E.A.T research environment and the Wenner-Gren Foundations.
P. T. H. P. is supported by the research program of the Netherlands Organization for Scientific Research (NWO).
H. Kumar thanks the LSSTC Data Science Fellowship Program, which is funded by LSSTC, NSF Cybertraining Grant \#1829740, the Brinson Foundation, and the Moore Foundation; his participation in the program has benefited this work.
S.McB. and J.M. acknowledge support from Science Foundation Ireland under grant number 17/CDA/4723. RD acknowledges support from the Irish Research Council (IRC) under grant GOIPG/2019/2033.
P.C. acknowledges support from the Department of Science and Technology via Swarana Jayanti Fellowship award (file no.DST/SJF/PSA-01/2014-15). We thank the staff of the GMRT that made these observations possible. GMRT is run by the National Centre for Radio Astrophysics of the Tata Institute of Fundamental Research.
We thank D. Bhattacharya, A. Vibhute, and V. Shenoy for help with CZTI analysis.

 \item[Competing Interests] The authors declare that they have no
competing financial interests.
\item[Contributions] TA and LPS were the primary authors of the manuscript. MMK is the PI of GROWTH and the ZTF EM-GW program, and SBC is PI of the SGRB program. MC, SA, IA, and MA support development of the GROWTH ToO Marshal and associated program. HK and CF led the reductions of the Gemini data. EB led analysis of the \textit{Fermi} gamma-ray data. GR, VC, TD, and PTHP contributed to the afterglow, \ac{KN}, and \ac{SN} modeling. RD and JM were the GBM burst advocates and provided gamma-ray analysis. DSS, DF, KH, AR and AT performed IPN and Konus analyses. AJCT, AV and SBP provided the GTC spectrum. KD performed the WIRC data reduction. PC and SP provided GMRT data. PG, SD and ET provided the LDT data.EH performed galaxy and SED fitting. SI and VB performed the Astrosat analyses. CC contributed to the GROWTH Marshal. BB, AG, DP, AYQH, VK, EK RS, SR, AS, RS contributed to candidate scanning, vetting, and classification. EB, DAD, MJG, RRL, SRK, FJM, AM, PR, BR, DLS, RS, MS, and RW are ZTF builders. All authors contributed to edits to the manuscript.

\item[Correspondence] Correspondence and requests for materials
should be addressed to Tom\'as Ahumada~(email: tahumada@astro.umd.edu).
\end{addendum}

\clearpage
\newpage

\begin{methods}

\section{Discovery} \label{sec:discovery}

\subsection{Background.}

The traditional paradigm of classifying a GRB based only on its gamma-ray properties is debated\citeNew{ZhCh2008,BrNa2013}. Although the two types of progenitors broadly map to the time duration of the signals, there is no clear boundary in the bimodal distribution. Some \acp{SGRB} show afterglow and host properties akin to \acp{LGRB}, e.g. Ref.~\cite{antonelli2009grb,levesque2010grb}; some \acp{LGRB} show no evidence for collapsars to deep limits akin to \acp{SGRB}, e.g. Ref.~\cite{yang2015possible,fynbo2006no}. 

In the mid 2000s, \ac{SGRB}-related breakthroughs triggered by the \textit{Neil Gehrels Swift Observatory}\cite{swift} and HETE-II\cite{ricker2003gamma} included detections of \acp{SGRB} X-ray afterglows, identification of likely host galaxies\cite{2005Natur.437..851G}, and the first \ac{SGRB} optical afterglow\cite{2005Natur.437..859H,villasenor2005discovery}. Yet, in the first decade of optical follow-up of \emph{Swift} \acp{SGRB}, only about 30 optical/NIR afterglows were detected\cite{berger2014,Fong15} and even fewer had associated redshifts.
On 2017 August 17, the joint detection of the \ac{BNS} merger (GW170817) in \acp{GW} with LIGO and Virgo\cite{2017PhRvL.119p1101A}, and in gamma-rays by \emph{Fermi} \ac{GBM}\cite{Meegan2009} and INTEGRAL\cite{2017ApJ...848L..13A,goldstein2017ordinary,savchenko2017integral}, unequivocally confirmed \ac{BNS} mergers as at least one of the mechanisms that can produce a \ac{SGRB}. The merger illuminated the entire electromagnetic spectrum\cite{2017ApJ...848L..12A} and the optical/NIR emission provided robust evidence of a radioactively powered \ac{KN}\cite{KaNa2017,2017Natur.551...64A,2017ApJ...848L..27T,2017ApJ...848L..19C,2017ApJ...848L..17C,2017Sci...358.1570D,2017Natur.551...80K,2017Natur.551...67P,2017Natur.551...75S,2017Natur.551...71T}.

The bulk of our knowledge about \acp{SGRB} comes from the well-studied \emph{Swift} \ac{SGRB} afterglow sample; however, the low \emph{Swift} local rate of \acp{SGRB}\cite{dichiara2020short} has hampered the detection of a GW170817-like event. On the other hand, the \emph{Fermi} \ac{GBM} is arguably
the most prolific engine of discovery for \emph{compact binary mergers}, as it detects $\approx$\,1 \ac{SGRB} each week, which is four times the rate of \emph{Swift}\cite{2011ApJS..195....2S,von_Kienlin_2020}, and comparable to the rate of the LIGO and Virgo detectors when in observing mode ($\approx$\,1.5 per week during O3a\cite{LIGO_O3a}). Unfortunately, there is a relative paucity of electromagnetic observations of \emph{Fermi} \acp{GRB} and LIGO/Virgo \ac{GW} events alike because it is challenging to pinpoint them within their positional uncertainties of tens to thousands of square degrees\cite{NissankeLocalization,aasi2016prospects,abbott2018prospects}.

In order to better understand the phenomenology of \acp{GRB} and compact binary mergers, wide \ac{FOV} optical instruments have looked for \emph{Fermi} \acp{GRB} and LIGO/Virgo \ac{GW} counterparts. In 2013, the \ac{PTF}\cite{2009PASP..121.1395L} used a 7\,deg$^2$ camera on the \ac{P48} to discover the first optical afterglow of a \ac{GRB} based \emph{solely} on a \emph{Fermi} \ac{GBM} localization\cite{2013ApJ...776L..34S} and subsequently found afterglows of 7 other \acp{LGRB}\cite{SiKa2015}. Now, the \acl{ZTF} (ZTF\cite{bellm2018zwicky,graham2019zwicky,MaLa2018}), a 47\,deg$^2$ camera mounted at the \ac{P48} telescope, has enabled searches an order of magnitude faster in areal and volumetric survey speed and has been used 
to search the coarse error regions of \emph{Fermi} \ac{GBM} \acp{SGRB}\cite{CoAh19sgrb} and LIGO/Virgo events\cite{coughlin2019growth,andreoni2019growth,kasliwal2020kilonova,anand2020nsbh}. Both \ac{PTF} and \ac{ZTF} have also discovered afterglow-like transients with optical, X-ray, and radio emission, but no gamma-ray counterpart\cite{cenko2013discovery,cenko2015iptf14yb,ho2020ztf20aajnksq}.

\subsection{Gamma-ray detections.}
\emph{Fermi} (\ac{GBM} trigger 620108997)\cite{GCN28284}, and the \ac{IPN} instruments -- \emph{AGILE}\cite{2020GCN.28289....1P}, \emph{INTEGRAL} (SPI-ACS), \emph{Mars-Odyssey} (HEND), Konus-\emph{Wind}~\cite{Ridnaia_2020_GCN28294}, and AstroSat\cite{Astrosatgcn} detected the burst.
The \ac{GBM} localization calculated by the ground software\cite{Connaughton2015, Goldstein2020} is RA = 4.7, Dec = 35.3 (J2000 degrees, equivalent to J2000 00h 18m, 35d 17') with a statistical uncertainty of 1.7 degrees\cite{GCN28284}. The BALROG localization is consistent with this\cite{balrog}.
The final \ac{GBM} localization of the GRB was available a few minutes after the trigger, and it enclosed 339 (63) deg$^2$ within the  95\% (50\%) credible region.

The light curve shows a bright GRB featuring three distinct peaks in the 50--300 keV energy range, with with a \ac{SNR} of $\sim$100 and a $T_{90}$ duration of 1.14 $\pm$ 0.13 s. 
This \ac{GRB} is within the top 36\% in terms of fluence (in the 50--300 keV energy band as measured over $T_{100}$) from bursts recorded in the \emph{Fermi} \ac{GBM} catalog\cite{von_Kienlin_2020}. Spectral analysis was performed using the RMFIT 4.4.2 light curve and spectral analysis software\footnote{Available at the \emph{Fermi} Science Support Center: \url{https://fermi.gsfc.nasa.gov/ssc/data/p7rep/analysis/rmfit/}} in the interval of $T_0$ to $T_{0}$ +1.152\,s encompassing the $T_{90}$ start time and duration. The detector selection for the analysis consisted of the brightest NaI detectors with angle $<$ 60$^{\circ}$ (N6, N7, N8, N9, NB11) and BGO detector B1. NaI and BGO detector data from $\sim$8\,keV to $\sim$900\,keV and $\sim$300\,keV to $\sim$40\,MeV respectively was used for the analysis. The Band Model\cite{Band1993}, a standard model for \ac{GRB} analysis\cite{Gruber2014}, provided best fit with parameters consistent with the initial analysis\cite{GCN28287}, where $E_{peak} = 88.9 \pm 3.2$ keV, $\alpha$ = $-0.26 \pm 0.07$, and $\beta = -2.4 \pm 0.1$. The peak flux in the 64 ms time (10 keV - 1 MeV), measured from $T_{0}$ +1.152\,s, is  $64.3  \pm  2.1$ ph cm$^{-2}$ s$^{-1}$. The fluence over $T_{90}$  (10 keV to 1 MeV) is $42.6 \pm 0.2 \times 10^{-7}$ erg cm$^{-2}$ (the fit merit of the spectral fit is 1.09).

A time resolved spectral analysis was performed from $T_{0}$ to $T_{0}$ +1.152\,s. The data was binned by \ac{SNR}, with at least 20$\sigma$ for the \ac{SNR} in each bin. For 6 of the bins, the best-fit parameters of the Band Model are displayed in \ref{tab:timeres_gbm}. The parameters for the Band model in the initial time interval were poorly constrained so a cutoff power law (Comptonised) model\cite{yu2016} was fitted to this interval. The Band Model was primarily used as this was the best constrained model for the time integrated analysis. The Castor C-stat statistic\cite{Guiriec2010} was used for its robustness when dealing with low count statistics, while the background was modeled as a first order polynomial over all qualifying detectors, as discussed above, using TTE data products. Spectral evolution of the $E_{peak}$ parameter is observed over the duration of the burst, and is found to exhibit hardness-intensity tracking behavior\cite{Guiriec2010,yu2016}.

The distribution of GRB $T_{90}$ durations demonstrates a bimodality that is best fit with double log-normal components. This suggests two distinct but overlapping progenitor probabilities, and by extension, classifications for short and long GRBs. Using only the $T_{90}$, and the parameters derived for the \textit{Fermi} \ac{GBM} distribution of bursts in Ref.\cite{BrNa2013}, we determine a \ac{SGRB} probability of $64.94\%^{11.59}_{-11.30}$ for GRB\,200826A.%

According to BATSE\cite{1993ApJ...413L.101K}, the overlap of GRB $T_{90}$ duration distributions occurs at $\sim$\,2\,s. In the past, GRB hardness ratios, quantifiable for the majority of GRBs in the absence of location information, have been used as a measure for the spectral hardness.
Hardness ratio is presented for the GBM catalogue events in Ref.~\cite{von_Kienlin_2020}, where the data shows an anti-correlation between hardness and duration; however, there is a large scatter in the data.  We derive a $T_{90}$ value of 4.2\,s as the duration where an event has equal probability of being in the short or long class\cite{von_Kienlin_2020} based on 2353 GRBs. We choose the $E_{\rm peak}$ parameter instead of hardness for separating LGRB and SGRB populations as it does not require set energy bands for a comparison to be made\cite{Goldstein_2010}. In Figure~\ref{fig:gamma-rays} we presents a log-log plot of the $T_{90}$ and $E_{peak}$ parameter, for which we fit two log-normal distributions, one for each class, and the probability that an event is an SGRB is translated into a colour. From this analysis, the probability that GRB\,200826A belongs to the short class is 74\% \cite{BurnsInPrep}. 

GRB~200826A triggered Konus-\emph{Wind} (KW) at $T_{0\mathrm{KW}}$ = 16195.106~s~UT (04:29:55.106). The propagation delay from Earth to Wind is 2.540\,s for this GRB; 
correcting for this factor, the KW trigger time corresponds to the Earth-crossing time 16192.566~s~UT (04:29:52.566).
The burst light curve shows a multi-peaked pulse which starts at about $T_{0\mathrm{KW}}-0.10$~s and has $T_{100}=0.97$~s, $T_{90}=0.67_{-0.03}^{+0.13}$~s, and $T_{50}=0.28_{-0.03}^{+0.04}$~s measured in the 80--1300~keV band.
Considering only the duration, the $T_{90}$ and $T_{50}$ of GRB~200826A are consistent with the short GRB population in the KW sample. The $T_{90}$ and $T_{50}$ durations at which a KW-detected GRB has an equal probability of being short- or long-duration
are 2~s and 0.7~s, respectively~\cite{Svinkin2019JPhCS1400}.
The spectral lag between the 20--80~keV and 80--330~keV 16~ms light curves is $30 \pm 11$~ms,
consistent with the bulk of KW short GRBs~\cite{Svinkin2016ApJS22410}.


During the burst, KW measured five spectra in the 20~keV--10~MeV band. 
The first four with 64~ms accumulation time cover the
interval from $T_{0\mathrm{KW}}$ to $T_{0\mathrm{KW}}+0.256$~s
and the fifth, from $T_{0\mathrm{KW}}+0.256$~s to $T_{0\mathrm{KW}}+8.448$~s.
The time-averaged spectrum of the burst (measured from $T_{0\mathrm{KW}}$~s to $T_{0\mathrm{KW}}+8.448$~s)
is best fit in the 20~keV--2~MeV range by the Band function with 
a low-energy photon index $\alpha = 1.26_{-1.12}^{+1.91}$,
a high-energy photon index $\beta = -2.32_{-0.15}^{+0.12}$, and
a spectrum peak energy $E_{peak} = 67_{-11}^{+13}$~keV
($\chi^2/\mathrm{d.o.f.} = 50/59$).
The burst
had a fluence of $4.60_{-0.60}^{+0.71} \times 10^{-6}$ erg~cm$^{-2}$,
and a 16-ms peak flux, measured from $T_{0\mathrm{KW}}+0.544$~s,
of $9.81_{-1.64}^{+1.83} \times 10^{-6}$~erg~cm$^{-2}$~s$^{-1}$
(both in the 20~keV--10~MeV energy range).
Using the spectroscopic redshift $z = 0.7481$\cite{GCN28319}
we have estimated the following rest-frame parameters:
the isotropic energy release $E_{iso}$ is $7.17_{-0.94}^{+1.11} \times 10^{51}$~erg,
the peak luminosity $L_{iso}$ is $2.67_{-0.45}^{+0.50}\times10^{52}$~erg~s$^{-1}$,
and the rest-frame peak energy of the time-integrated spectrum,
$E_{peak,z}$ is $117_{-19}^{+23}$~keV.

With these values, GRB~200826A is within the softest $\sim 1$\% of KW short GRBs 
in terms of the observed $E_{peak,z}$ and is within the $1 \sigma$ prediction band of
both the `Amati'  and `Yonetoku' relations based on 315 long/soft (Type~II)
GRBs with known z\cite{Tsvetkova2017ApJ850161, amati2002intrinsic,yonetoku2004gamma}.
Furthermore, in the $E_\mathrm{iso}$-$E_\mathrm{peak,z}$ plane, GRB~200826A is inconsistent with the short-hard (Type~I) GRB population (see \ref{fig:grb_lc}).
Thus, despite the short duration, the KW parameters of GRB\,200826A imply that it belongs to the
long/soft GRB population.

In addition to the IPN detections, the Cadmium Zinc Telluride Imager (CZTI\cite{Bhaleraoetal2017}) on board AstroSat also detected the burst. We reanalysed the data, combining data from all four quadrants to create 20--200~keV light curves with 0.05~s, 0.1~s and 0.2~s bins. All light curves show a single pulse,  with hints of sub-structure in during the rise in the smaller time bins (see \ref{fig:grb_lc}). We process the 0.05~s light curves with the CIFT pipeline\cite{sharma2020search}, which incorporates better data analysis as compared to the quick-look pipeline and produces more robust results. Our reanalysis yields a peak time of UT 04:29:52.95 - consistent with the Fermi peak. The new value of $T_{90}$ is $0.94^{+0.72}_{-0.18}$ s, significantly shorter than the quick-look values reported in Ref. \cite{Astrosatgcn}, but consistent with Fermi and Konus-Wind.

\subsection{Optical.}
We ingested the \ac{GBM} localization map into the GROWTH \ac{ToO} marshal, an interactive tool design to plan and schedule \ac{ToO} observations for \ac{ZTF}\cite{CoAh19sgrb}. The observation plan generated by the \ac{ToO} marshal relies on \texttt{gwemopt}\cite{CoTo2018,CoAn2019,AlCo2020}, a code that optimizes the telescope scheduling process for skymaps with a healpix format, like the \emph{Fermi}-\ac{GBM} maps. The \texttt{gwemopt} procedure involves slicing the skymap into predefined tiles of the size and shape of the \ac{ZTF} field-of-view, determining which fields have the highest enclosed probability, and optimizing observations based on airmass and visibility windows. For this purpose, we used a modified version of the ``greedy" algorithm described in Ref.~\cite{rana2017enhanced}, implemented  within \texttt{gwemopt}.
The resulting optimized plan for the optical follow-up of GRB\,200826A consisted of four primary \ac{ZTF} fields and one secondary field\footnote{The \ac{ZTF} secondary fields are strategically located to cover the chip-gaps of the fields in the primary grid}. The fields were observed in the $r$- and $g$-band for 300\,s each, starting 4.9 hours after the \ac{GBM} detection. The observing plan for the first night covered 186 deg$^2$, corresponding to 77\% of the \ac{GBM} region (see Fig.\ref{fig:1Discovery}). 
Once the Konus-\emph{Wind} data
became available on ground at about 17:55 UT, the 288~arcmin$^2$ \ac{IPN}
error box was derived using Konus, \ac{GBM}, and HEND data, which allowed
the X-ray Telescope \acused{XRT}(XRT\cite{xrt}) on board \textit{Swift} to initiate a \ac{ToO} at about 20:45~UT. The \ac{IPN} box was published
later at about 21:30~UT\cite{GCN28291}.

Reference images of the fields are then subtracted from the ZTF \ac{ToO} observations and any high significance difference ($>$5$\sigma$) generates an \textit{alert}\cite{MaLa2018,2019PASP..131a8001P} that contains relevant information about the transient. We queried the stream of alerts using three different tools: the GROWTH marshal\cite{Kasliwal2018}, the Kowalski infrastructure\cite{kowalski}\footnote{\url{https://github.com/dmitryduev/kowalski}} and AMPEL\cite{ampel,2018PASP..130g5002S,Stein2020TDE,robert_stein_2020_4048336}\footnote{\url{https://github.com/AmpelProject}}. Our filtering scheme has been described in previous \ac{SGRB} and \ac{GW} searches\cite{CoAh19sgrb,kasliwal2020kilonova}, but we summarize the main points here. We aim to identify sources that 1) are spatially coincident with the skymap, 2) are detected only after the \ac{GBM} trigger, 3) are far from known bright sources, 4) are spatially distinct from \acl{PS1} (PS1\cite{2016arXiv161205560C}) stars (based on Ref.~\cite{TaMi2018}), 5) have at least two detections separated by at least 15\,minutes to avoid moving objects, 6) have a real-bogus score (RB\cite{mahabal2019machine,kowalski}) greater than 0.15, and 7) showed an increase in their flux relative to the reference image. 
For each candidates passing these filters, we visually inspected the light curve, cross-matched with the PS1-DR2 catalog to check for previous activity, and cross-matched against the \acl{WISE} (WISE\cite{2013wise.rept....1C}) catalog to determine whether it was consistent with an \ac{AGN} based on its position in the WISE color space\cite{WrEi2010,StAs2012,assef2018}.
Additionally, we searched the \ac{MPC} to ensure that our potential counterparts were not consistent with known solar system objects. Forced photometry\cite{Yao2019} was performed using data from the 10 nights previous to the trigger, to remove young \acp{SN} from the sample. A final quality check was done by querying for alerts around each candidate, and rejecting transients that have multiple alerts within a radius of 15'' as these alerts could suggest artifacts, ghosts, or slow-moving objects near the stationary points\cite{Jedicke2016}. 

After the first night of observations 28195 alerts were generated in the region and 14 sources passed every stage of our filtering criteria\cite{GCN28293}. The \ac{SGRB} afterglow, ZTF20abwysqy\cite{GCN28295}%
%
%
, was not in our first selection as there was a ZTF alert 11'' NE; however, it passed all the other filtering criteria. 
Two of the \ac{ZTF} \ac{ToO} fields triggered during the first night covered $>99\%$ of the triangulated IPN region. However, none of the candidates reported in Ref.~\cite{GCN28293} fell there. 
For our second night of observations, we scheduled 600\,s observations in $r$- and $g$-band and queried the database.

\subsection{Probability of Chance Coincidence.}

We roughly calculate the probability of chance coincidence ($p$) of this optical transient to be independent of GRB\,200826A. For this we follow Ref. \cite{stalder2017observations} and use Poissonian statistics to derive the probability for one or more events to be randomly coincident:
\begin{equation*}
    p =  1 - e^{-\lambda} 
\end{equation*}
where $\lambda$ is the product of three different values, $\lambda = {\displaystyle \prod_{i=1}^{3} r_{i}}$. For this study, we consider $r_1$ to be the time window between the last \ac{ZTF} non-detection (0.72 days before the trigger time) and the time of the detection (0.21 days, see \ref{table:observations_afterglow}), resulting in $r_1 = 0.93$ days. The parameter $r_2$ is the rate of \textit{Fermi} \acp{GRB}, which from the latest \ac{GBM} catalog\cite{von_Kienlin_2020} gives $\sim 0.65$ bursts per day. Finally $r_3$ is the ratio between the total IPN area (288 arcmin$^2$) and the sky.  
This derivation gives a $p=1.16e-6$, and allows us to rule out a random association between the \ac{GRB} and the afterglow at the 4.87$\sigma$ level.

\section{Follow-up} \label{sec:followup}

The summary of the follow-up results can be found in \ref{table:observations_afterglow},\ref{table:observations_xrt}, and \ref{table:observations_radio}.

\subsection{Optical/NIR.}

Here we present our optical and \ac{NIR} follow-up results. In addition to our observations, MASTER\cite{lipunov2005master} and Kitab follow-up with a clear filter\cite{lipunov2020fermi,GCN28306} led to upper limits of 18.3 mag and 20.4 mag respectively.

\emph{Las Cumbres Observatory.} We performed follow-up of ZTF20abwysqy with the Spectral camera mounted at the 2-m Faulkes Telescope North (FTN), located at Haleakala Observatory.  Starting on 2020-08-28 11:32:26 UT we acquired three sets of images with 300s exposures each in $g$- and $r$-bands 
through the LCO observation portal\footnote{\url{https://observe.lco.global/}}. 
Furthermore, we obtained a second (reference) epoch in $g$- and $r$-bands 
on 2020-09-02 UT.  Our images were reduced by an automatic subtraction pipeline. Our pipeline retrieves images from LCO, stacks them, extracts sources from the image using \texttt{Source Extractor}\cite{sextractor} and performs photometric calibration of sources using the PS1 catalog.  Then, the pipeline performs image subtraction using the High Order Transform of Psf ANd Template Subtraction code (HOTPANTS\cite{hotpants}) to subtract a PSF-scaled reference image aligned using \texttt{SCAMP}\cite{scamp}. Our photometry measurements reported in $g$- and $r$-band
were determined after host subtraction using the LCO reference images in the same filters acquired on 2020-09-05.

\emph{Hale Telescope.} We obtained two epochs of dithered J-band imaging with the Wide Field Infrared Camera (WIRC\cite{Wilson2003}) on the Palomar Hale 200-in (P200) telescope on 2020-08-28 UT and on 2020-09-04 UT, spending an hour integrating on target during each epoch. We reduced the images using the image reduction pipeline described in Ref.~\cite{DeHa20}.  Images were aligned and stacked using \texttt{SWarp}\cite{swarp} and calibrated against the 2MASS catalog. Image subtraction between the two epochs was performed using the method described in Ref.~\cite{De2020} to derive flux measurements and its uncertainty for the first epoch.

\emph{Gran Telescopio Canarias.}  The 10.4m 
Gran Telescopio Canarias (GTC) (+OSIRIS) obtained spectroscopic observations starting on August 30, 04:30 UT. Two 1200s exposures were gathered with the R1000B grism and one 1200s exposure was obtained with the R2500I grism, in order to cover the entire 3700--10000 \AA \ \rm range. The slit was placed covering the position of the potential host galaxy. Standard routines from the Image Reduction and Analysis Facility (IRAF) were used to reduce the data.

\emph{Lowell Discovery Telescope.} 
We used the Large Monolithic Imager (LMI\cite{massey2013big}) mounted on the 4.3m Lowell Discovery Telescope (LDT) to observe the optical transient ZTF20abwysqy on three different nights: August 29, September 13, and September 19 (3.2, 18.1 and 24.2 days after the GRB trigger). Observations were conducted with an average airmass of 1.0 while the seeing varied from 1.1\arcsec\ for the first night to 1.4\arcsec\ and 1.6\arcsec\ for the second and third night, respectively. 
We took 8 exposures of 180~s in the $r$-band on August 29. Images were taken with different filters on September 13: 5 exposures of 180~s in $u$-band, 4 exposures of 180~s in $g$-band, 6 exposures of 180~s in $i$-band
and 6 exposures of 180~s in z-band. Finally, 10 exposure of 150 s in the $i$-band and $r$-band were taken during the last night of observations on September 19.
We used standard procedures to perform bias and flat-field correction. The astrometry was calibrated against the SDSS catalog (release DR16\cite{dr16sdss}) and frames were aligned using \texttt{SCAMP} and stacked with \texttt{SWarp}. After stacking the images, we extracted sources using \texttt{Source Extractor} and the magnitudes were calibrated against 45 \ac{PS1} stars in the field on average.
We used HOTPANTS to perform image subtraction between the first and third epochs and found a source in the $r$-band at the location of ZTF20abwysky with a magnitude of 24.46$\pm$0.12 mag (see \ref{table:observations_afterglow}). We determine the magnitudes of the host galaxy using the second epoch of observations. To verify this result, we used the 80 days $i$-band \ac{GMOS}-North (see description in the paragraph below) as a reference and the HOTPANTS subtraction shows a source with $r$-band magnitude of 24.76$\pm$0.23 mag, consistent with the result using the LDT reference.



\emph{Gemini Observatory.}
We acquired images of the transient location on September 23, October 10, and November 7 corresponding to 28.28, 46.15 and 80.23 days after the \ac{GRB} trigger respectively, which we denote as epoch 1, 2, and 3.
We used \ac{GMOS}-North, mounted on the Gemini North 8-meter telescope on
Mauna Kea, under the approved \ac{DDT} proposal DD-104 (P.I.: L. Singer). 
The host galaxy coordinates, in the Gemini images, are $\alpha=00^d27^m08.5557^s$, $\delta= +34^d01^m38.634^s$.

The first set of observations (epoch 1) was scheduled to be closest in time to a SN1998bw-like peak without suffering from moon illumination. All three sets of observations consisted of 14 200\,s $r$- and $i$-band exposures, with a position angle of 45$^\circ$ to avoid blooming from neighboring bright stars. The average airmass for the observations was $\sim$\,1.0 and the seeing was stable throughout the three epochs, at 0.7\arcsec, 0.5\arcsec, and 0.85\arcsec. The images were later reduced using DRAGONS\footnote{\url{https://dragons.readthedocs.io/}}\cite{2019ASPC..523..321L}, a Python-based data reduction platform provided by the Gemini Observatory. 



We extracted sources using \texttt{Source Extractor} and determined a photometric zero point using 23 \ac{PS1} stars in the field. Using the ZOGY\cite{ZaOf2016} algorithm-based python pipeline, we performed image differencing on Gemini data. The pipeline makes use of \texttt{Source Extractor}, \texttt{PSFEx}\cite{psfex}, \texttt{SWarp}, \texttt{SCAMP}, and \texttt{PyZOGY}\cite{Guevel2017Pyzogy} to perform image subtraction. We took a 1300 pix $\times$ 1300 pix sized cutout for both the science and reference images, centered at the ZTF20abzwysqy position in the image in order to achieve good subtraction quality, as the images were affected by background variation. Sources were  extracted using \texttt{Source Extractor} from both science and reference images with a 5-$\sigma$ detection threshold. The resulting catalogues were fed to \texttt{SCAMP}, which calculates and corrects for the astrometric errors in both science and reference catalogues with the help of a Gaia data release 2 catalogue\cite{2018A&A...616A...1G} for the same field. Using \texttt{SWarp}, images are then re-sampled to subtract the background and the \ac{PSF} of the images. \texttt{Source Extractor}-generated weight maps were used as input to the \texttt{SWarp} to generate variances, which when added with the images' Poisson noise in quadrature, which results in the rms image. We select good sources from \texttt{Source Extractor} catalogues based on their \ac{SNR}, \ac{FWHM} and \texttt{Source Extractor} flag values. Sources which raised a \texttt{Source Extractor} flag were discarded. These sources were used to match the flux level of the images. We used \texttt{PSFEx} to extract the PSF model of the re-sampled images with the \texttt{Source Extractor} catalogue as input to \texttt{PSFEx}. The rms images, re-sampled images, \ac{PSF} models and astrometric uncertainty are used as input to \texttt{PyZOGY}, which generate the difference image and a corrected score image as final products\cite{ZaOf2016}. A source is detected with a ZOGY corrected score of~4 in the $i$-band, at the location of the transient. With ZOGY, we derive an $i$-band magnitude of 25.49$\pm$0.12 mag for epoch 1. No source is detected in the r-band of epoch 1 nor in any band in epoch 2.
 
We have confirmed the ZOGY-based results using an independent image subtraction pipeline, FPipe\cite{FrSo2016}, which is based on empirically measuring the \acp{PSF} of the science and reference images and matching them using the common \ac{PSF} method (CPM\cite{2008ApJ...680..550G}). We detect the source with an $i$-band brightness of 25.45$\pm$15 mag in epoch 1. The results agree within uncertainties with the ZOGY-based subtractions. We do not detect the source in the $r$-band up to a 5-$\sigma$ limit of $r$ $>$ 25.6\,mag for epoch 1. No source is detected in either filter during epoch 2 up to a 5$\sigma$ upper limit of $i$ $>$ 25.4 mag and $r$ $>$ 25.5 mag\cite{GCN29029} (see Figure \ref{fig:lightcurve}).

Therefore, using two independent image subtraction pipelines, we confirm the detection of a source in the $i$-band, at the location of the transient.


\subsection{UV/X-ray.}
\emph{Swift} began observing the IPN localization region of GRB\,200826A 0.7 days after the trigger. The last two observations were $\sim$4 ks and triggered as \ac{ToO} observations. The \ac{XRT} data was reduced by the the online reduction pipeline \footnote{\url{https://www.swift.ac.uk/xrt_curves/00021028/}}\cite{evans2007,evans2009}. As the hardness ratio remains constant within error bars, we assume a single absorbed power law spectrum, with a Galactic neutral hydrogen column\cite{willingale2013calibration} of to $6.02 \times 10^{20} $cm$^{-2}$.  We convert the count rates (see \ref{table:observations_xrt}) to flux density at an energy of 1 keV, using the parameters derived by the \emph{Swift} pipeline: a photon index of $\Gamma_X = 1.5^{+0.7}_{-0.5}$,  an intrinsic host absorption of $n_{H,int} = 6^{+32}_{-6} \times 10^{20} $cm$^{-2}$, and an unabsorbed counts-to-flux conversion factor of $4.27 \times 10^{-11}$ erg cm$^{-2}$ ct$^{-1}$. 

The first \ac{UVOT} observations of ZTF20abwysqy was $\sim$1.6 days after the burst and the detection in the white filter was associated to the underlying galaxy\cite{GCN28300}.

\subsection{Radio.}
For the afterglow modeling, we additionally use data reported by Ref.~\cite{GCN28302}. They measure a flux density of $\sim 40$ $\mu$Jy at a mean frequency of 6 GHz using the \ac{VLA}, 2.28 days after the trigger. We observed GRB\,200826A in band-5 (1050--1450 MHz) of the upgraded Giant Metrewave Radio Telescope (uGMRT\cite{gupta2017upgraded}) on 2020-09-09 15:41:30.5 UT and 2020-09-14 23:38:25.5 UT ($\sim$ 14.5 and $\sim$ 19.8 days after the burst respectively) under the approved \ac{DDT} proposal ddtC147 (P.I.: Poonam Chandra). The data was recorded with the GMRT Wideband Backend (GWB) correlator with a bandwidth of 400 MHz divided in 2048 channels, with the central frequency of 1250 MHz. The total on-source time was $\sim$75 mins with overheads of $\sim$30 mins. 3C48 was used as the flux and bandpass calibrator and J0029+349 was used as the phase calibrator. We used standard data reduction procedures in Common Astronomy Software Applications (\texttt{CASA})\cite{mcmullin2007casa} for analysing the data. The dead antennas were first flagged by manual inspection and the end channels were flagged due to low gain. The automatic flagging algorithms incorporated into the \texttt{CASA} task \emph{flagdata} were used to remove most of the Radio Frequency Interference (RFI). Any remaining corrupted data were then flagged manually. The calibrated data was then imaged and self-calibrated to get the final image. The synthesized beam for the final image was $\sim$4\arcsec$\times$2\arcsec. We found no evidence of radio emission at the GRB position on both days. The 3$\sigma$ upper limits were 48.6 $\mu$Jy/beam for the observations taken on 2020-09-09\cite{GCN28410} and 57.4 $\mu$Jy/beam for the observations taken on 2020-09-14.

 \section{The Host Galaxy} \label{sec:host} 
 
Our \ac{GTC} spectra of the host galaxy show strong [OII] and [OIII] features at $z=0.748$, in agreement with a previous report based on data from the Large Binocular Telescope Observatory (LBTO)\cite{GCN28319}.

We modeled the \ac{GTC} spectrum with the \ac{pPXF}\cite{Cappellari2017} to infer the properties of the host galaxy's stellar populations, resulting in a fit with $\chi^2_\nu = 1.011$. The \ac{pPXF} results confirmed the [OII] features at 3726\AA~and [OIII] lines at 4959\AA~and 5007\AA, indicators of recent star-formation and young, hot stars. The stellar population age derived from the weighted ages of the templates is 1.514$_{-0.85}^{+3.83}$ Gyr, with evidence for a younger $\sim$0.1 Gyr population (see Figure \ref{fig:galaxySED} and \ref{tab:fluxes}). 

Additionally, we corrected the magnitudes of the host galaxy (see \ref{table:observations_host}) using foreground extinction maps \cite{schlafly2011measuring} and fed them to \texttt{Prospector}\cite{prospector1}, to model the \ac{SED} of the host galaxy. \texttt{Prospector} uses \texttt{fsps}\cite{Conroy2009,Conroy2010} to generate multiple stellar populations, and fits the observed photometry to determine the formed mass of the galaxy, age, and intrinsic extinction among other parameters, using the WMAP9 cosmology\cite{Hinshaw2013} internally. We fitted the photometry to a galaxy using the Chabrier\cite{Chabrier2003} \ac{IMF}, the Calzetti\cite{Calzetti2000} extinction curves for the dust around old stars, and a star formation history (SFH) with the form of $t e^{-t/\tau}$. Our results using the nested sampling \texttt{dynesty}\cite{speagle2020dynesty} algorithm, gave a galactic stellar mass distribution of $M_{gal}= 4.64\pm 1.67\times10^9$\,$M_\odot$, a mass-weighted galactic age of $t_{gal} =1.08^{+1.28}_{-0.72}$\,Gyr, a metallicity of $log(Z/Z_\odot) = -1.06^{+0.67}_{-0.48}$, a dust extinction of $A_V=0.34^{+0.33}_{-0.22}$, and a \ac{SFR} of $ 4.01^{+41.87}_{-3.59}$ \(M_\odot\) yr$^{-1}$ (see the resulting \ac{SED} in \ref{fig:galaxySED} and the posterior probability distributions for the free parameters in \ref{fig:corner_sed}). We derived the \ac{SFR} and weighted-mass age  following similar studies on \ac{GRB} host galaxies\cite{nugent2020distant,paterson2020discovery,oconnor2020tale}, and the stellar mass from the mass fraction derived with \texttt{Prospector}.

\section{Modeling}\label{sec:afterglow} 

For the modeling of the multiwavelength emission, all optical and NIR observations were from difference images and were corrected for foreground extinction \cite{schlafly2011measuring}.

\subsection{The afterglow.}
In the standard synchrotron fireball model a power-law energy distribution characterized by the index $p$, $N(E) \propto E^{-p}$, results in a \ac{SED} described by a series of broken power laws\cite{sari98,granot02}. 
The frequencies at which the broadband \ac{SED} presents its breaks are the self-absorption frequency $\nu_a$, the synchrotron frequency $\nu_m$, and the cooling frequency $\nu_c$.

The temporal decline of both the optical and X-ray data during the first four days can be described by a single power law model which suggests the jet-break has not yet occurred. Our ZTF $g$-band observations are the most constraining in the optical therefore we use these to estimate the temporal decline rate $\alpha_\mathrm{o} = -1.05 \pm 0.13$. We use the 1 keV XRT data to find $\alpha_\mathrm{x} = -0.89 \pm 0.07$. The similar slopes between the optical and X-ray observations suggest the location of the cooling break frequency, $\nu_c$, lies beyond X-ray frequencies. Therefore, we estimate the spectral index at $\sim$1 day between the X-ray and optical as $\beta_{\mathrm{ox}} = -0.67 \pm 0.02$.

We now use $\alpha_\mathrm{o}$ and $\alpha_x$ to estimate the power law index of the electron energy distribution, $p$, and to determine the circumburst density profile. In a constant density (ISM-like) medium $\alpha_\mathrm{ISM} = \frac{3(1-p)}{4} \ (\nu < \nu_c)$, which gives $p_\mathrm{o} = 2.44 \pm 0.17$ and $p_\mathrm{x}= 2.17 \pm 0.10$\cite{Zhang2006}. In the wind-like scenario $\alpha_\mathrm{wind} = \frac{(1-3p)}{4}$, which gives $p_\mathrm{o} = 1.78 \pm 0.17$ and $p_\mathrm{x} = 1.50 \pm 0.10$.  The optical to X-ray spectral index $\beta_{\mathrm{ox}}$ gives an estimate for $p$ of $2.34 \pm 0.04$ ($\beta = \frac{1-p}{2}$). Theoretical studies of relativistic collisionless shocks predict $p \gtrsim 2$ and particularly $p\sim2.2$ in the ultra-relativistic limit~\cite{Sironi:2015aa}.  Given the low values of $p$ for the wind-like scenario we choose to assume an ISM-like density profile and $p\sim2.4$ throughout this work.


We note the possibility of a wind-like environment with $p < 2$ and a spectral break $\nu_c$ between the optical and X-rays.  Such a scenario would not change the predicted optical and NIR emission and it would be unable to account for the late GMOS $i$-band detection without an additional component.  However, this scenario may allow for a better fit of the afterglow X-ray evolution, which would decay at a rate shallower than the optical by $\Delta \alpha = 0.25$. We do not consider this scenario further because of the theoretically disfavored value of $p$. 

\subsection{Bayesian afterglow modeling.}
We now use Bayesian inference to analyze the X-ray, optical, NIR, and radio counterpart. We use two independent pipelines, one using Markov chain Monte Carlo (MCMC) based on the \texttt{EMCEE} Python package\cite{Foreman-Mackey2013} and one using nested sampling based on \texttt{PyMultinest}\cite{2014A&A...564A.125B}. The pipelines use different priors and implementations, but arrive at consistent results. Both utilize the \texttt{afterglowpy} python package\cite{Ryan2020} to estimate the physical parameters of the multi-wavelength afterglow. \texttt{Afterglowpy} is a public, open-source computational tool which models forward shock synchrotron emission from relativistic blast waves as a function of jet structure and viewing angle. Descriptions of the MCMC implementation may be found in Refs.~\cite{Troja2018,cunningham2020grb} and the nested sampling implementation in Ref.~\cite{DiCo2020}.  


For this work we assume Gaussian statistics for the optical and radio data, while assuming Poissonian statistics (via the C-statistic\cite{Cash:1979aa}) for the X-ray data due to low detector counts. The \texttt{Afterglowpy} model is parametrized by the isotropic kinetic energy, $E_{\mathrm{K,iso}}$; jet collimation angle, $\theta_c$; viewing angle, $\theta_v$; the circumburst constant density, $n$; the spectral slope of the electron distribution, $p$; the fraction of energy imparted to both the electrons, $\epsilon_e$, and to the magnetic field, $\epsilon_B$, by the shock. The redshift and luminosity distance of the source are held fixed.

Our modelling of the host indicates a small galaxy of stellar mass $\sim 5\times 10^9 M_\odot$ with moderate extinction $A_V=0.34^{+0.33}_{-0.22}$ (see \S \ref{sec:host} for details). In the MCMC implementation, we incorporate a Small Magellanic Cloud-like host extinction correction with total-to-selective extinction $R_V = 2.93$\cite{Pei:1992aa} implemented with the \texttt{dust-extinction} software package\cite{astropy}. We leave the color excess $E(B-V)$ as a free parameter, with a prior distribution computed from the $A_V$ posterior found from modelling the host galaxy. This posterior is similar to the distribution of host extinction values observed in \acp{LGRB} with well-sampled multiband photometry\cite{schady2015,littlejohn2015,zafar2018}. The nested sampling implementation performs no extinction correction.

Using a top hat model for the jet structure, we perform a search over the parameter space by allowing all the parameters  $E_{\mathrm{K,iso}}$, $\theta_c$, $\theta_v$, $n$, $p$, $\epsilon_e$, $\epsilon_B$, and $E(B-V)$ to vary with broad priors. We report only the MCMC results, although the nested sampling results are consistent. Given both the highly degenerate nature of the afterglow fitting and the low number of observations available, some of the parameters are not particularly constrained. One exception is the spectral slope of the electron distribution, for which we find $p = 2.4 \pm 0.04$, consistent with the analytical results. The uncertainty in the circumburst density, $n = 5.5^{+187.3}_{-5.4} \times 10^{-2}$\,cm$^{-3}$, includes typical ranges for both \acp{SGRB} and \acp{LGRB}. We find very little reddening from the host, $E(B-V) = 2.5^{+4.8}_{-2.3} \times 10^{-2}$.
The posterior probability distributions are shown in \ref{fig:corner_eps} and the parameter estimates are listed in \ref{tab:pars}.

\subsection{Bayesian model selection.}
In addition to an afterglow-only model, we consider an afterglow plus a \ac{KN} and an afterglow plus a \ac{SN}.

\textit{Kilonova.} We use \acp{SED} simulated by the multi-dimensional Monte Carlo radiative transfer code \texttt{POSSIS}\cite{Bulla2019}. The simulations are performed over a grid of \ac{KN} parameters: dynamical ejecta $ M^{\rm dyn}_{\rm ej}$, disk wind ejecta $M_{\rm ej}^{\mathrm{wind}}$, opening angle $\Phi$, and the observation angle $\Theta_{\rm{obs}}$ (see Ref.~\cite{DiCo2020} for details). We use Gaussian process regression \cite{Coughlin:2018miv,Coughlin:2019zqi} to interpolate the model, enabling rapid parameter inference.

\textit{Supernova.} The \ac{SN} model starts with a bicubic spline in time and frequency, $L_\nu^\mathrm{SN}(t, \nu)$, that interpolates a K-corrected SN1998bw template\cite{clocchiatti2011ultimate}. We apply to the template a scale factor, $k$, and a stretch factor, $s$, which are drawn from a bivariate normal distribution that is consistent with a historical sample of GRB-SNe\cite{cano2017,klose2019four}. The model for the observed flux density is
$$
F_\nu^\mathrm{SN} \left(t_\mathrm{obs}, \nu_\mathrm{obs}\right)
=
\frac{(1 + z) k}{4 \pi {d_\mathrm{L}}^2} L_\nu^\mathrm{SN}
\left(
\frac{t_\mathrm{obs}}{(1 + z) s},
(1 + z) \nu_\mathrm{obs}
\right).
$$

We perform Bayesian model selection to determine which model best explains the data. We used the nested sampling pipeline to calculate the Bayesian evidence for each of the three models. The Bayes factor, or the ratio of the evidences, between the afterglow-plus-\ac{KN} model and the afterglow-only model is $\sim1$, indicating that neither model is strongly favored over the other, because the \ac{KN} contributes negligible flux compared to the afterglow at the time of the GMOS observation. The Bayes factor between the afterglow-plus-\ac{SN} model and the afterglow-only model is $\sim10^{5.5}$, strongly favoring the presence of a \ac{SN}.

To better understand the source of discriminating power between models, we carry out a posterior predictive check. Here, we sample the posterior while excluding the GMOS $i$-band data point and then predict its value using the rest of the data. The posterior predictive distribution of the AB magnitude for the GMOS $i$-band detection both with and without the inclusion of a \ac{SN} contribution is shown in \ref{fig:posterior_prediction_afterglow_supernova}. The fit with the \ac{SN} is consistent with the observation, while the fit without the \ac{SN} is inconsistent at the $\sim 5\sigma$ level. Therefore, inclusion of the GMOS $i$-band data point requires a \ac{SN} component, confirming a collapsar origin.

We compared our GMOS-N detection against extinction-corrected $i$-band fluxes of three GRB-\ac{KN} candidates found in the literature\cite{2017Sci...358.1559K}; using photometry from GRB\,130603\cite{berger2013r} and GRB\,160821\cite{troja2019afterglow,2017ApJ...843L..34K}, and the compiled light curve of AT2017gfo\cite{2017Sci...358.1556C,ViGu2017}, we correct to the redshift of GRB\,200826A. We fit each light curve to the best-fit 2D model of GW170817\cite{DiCo2020} from \texttt{POSSIS} using SNCosmo\cite{2016ascl.soft11017B} and extract the corresponding $i$-band magnitude at a rest-frame time of 16\,days. At $z=0.748$, both AT2017gfo and GRB\,160821 would be at M$\approx-10$\,mag, $\sim$8\,mags fainter than our detection. GRB\,130603B does not have enough late-time detections at the same phase for comparison.

Our detection of a source with an extinction-corrected absolute magnitude in the $i$-band with $M_i = -18.0$ is consistent with the population of collapsars associated with \acp{LGRB}. A typical \ac{SN} Ic reaches its peak magnitude 10 to 20 days post-burst, at $M_{B} = -17.66 \pm 1.18$ mag \cite{richardson2014absolute}. 
Fig.~\ref{fig:lightcurve} shows K-corrected light curves for three well-sampled GRB-SNe: SN1998bw, one of the brightest; SN2006aj and SN2010bh, among the faintest.

\end{methods}

\renewcommand{\thefigure}{Extended Data Figure \arabic{figure}}
\renewcommand{\figurename}{}
\setcounter{figure}{0}

\renewcommand{\thetable}{Extended Data Table \arabic{table}}
\renewcommand{\tablename}{}
\setcounter{table}{0}


\begin{figure*}
\begin{minipage}[b]{0.45\linewidth}
    \centering
    \includegraphics[width=\textwidth]{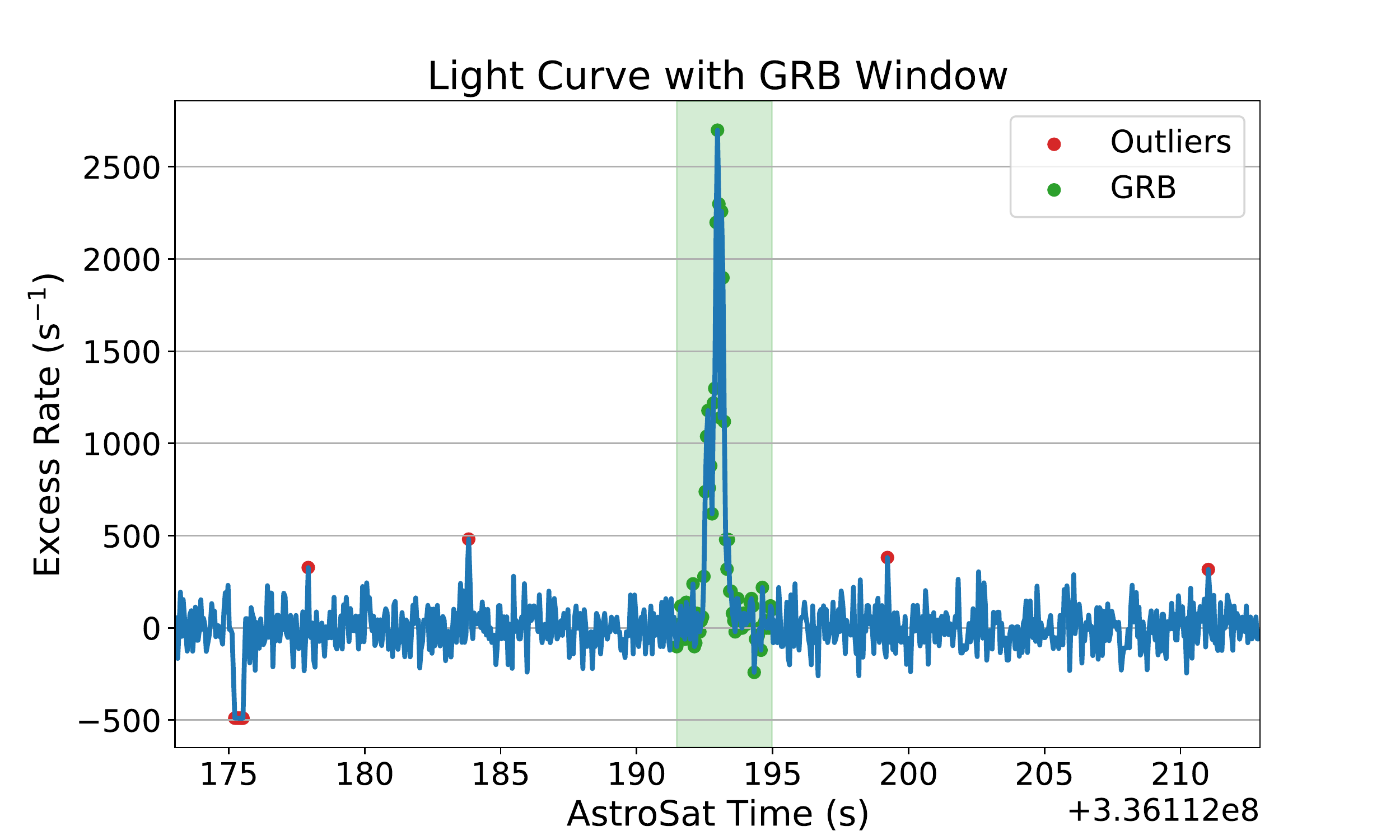}
    \includegraphics[width=\textwidth]{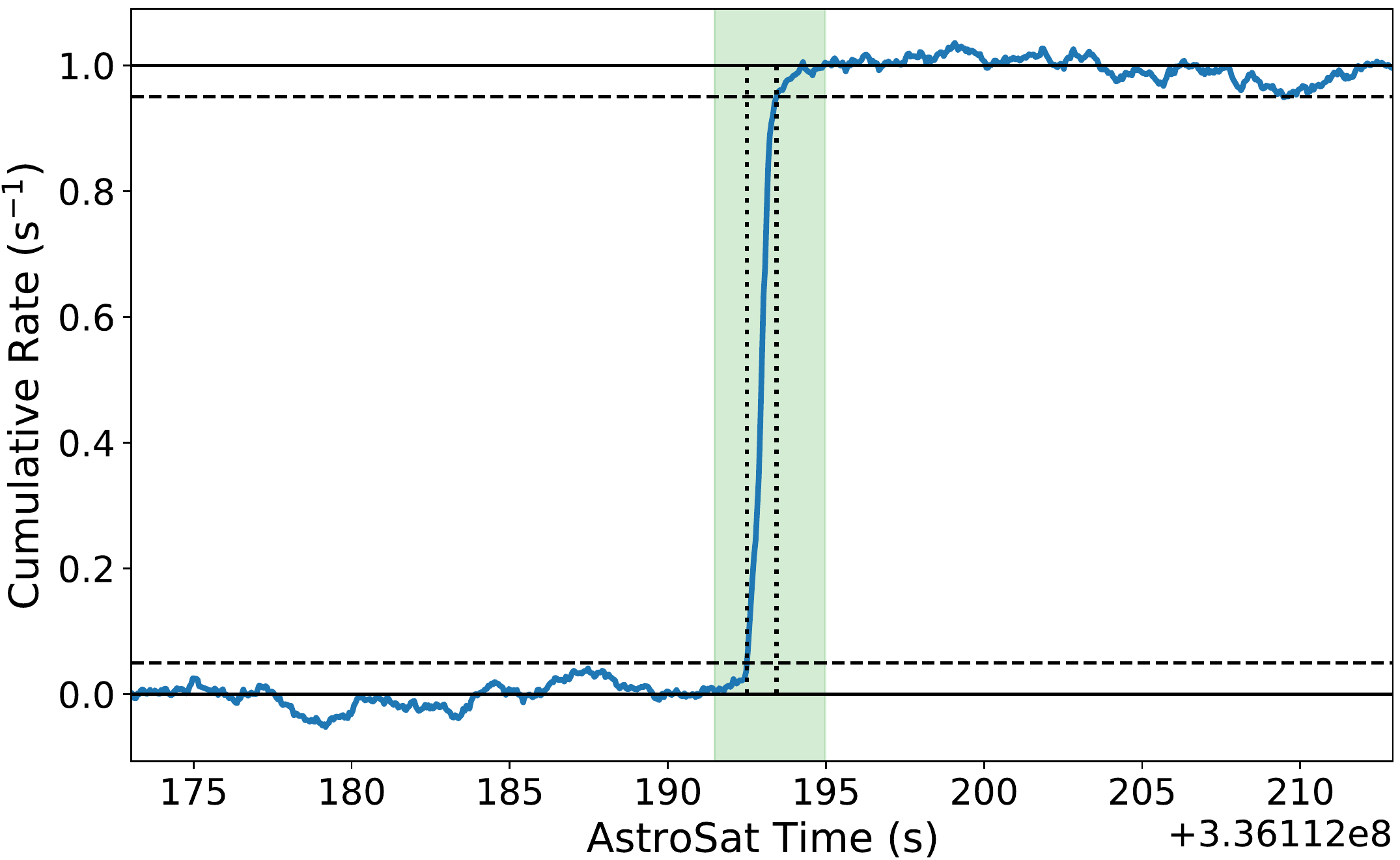}
  \end{minipage}
  \begin{minipage}[b]{0.65\linewidth}
    \centering
    \includegraphics[width=\textwidth]{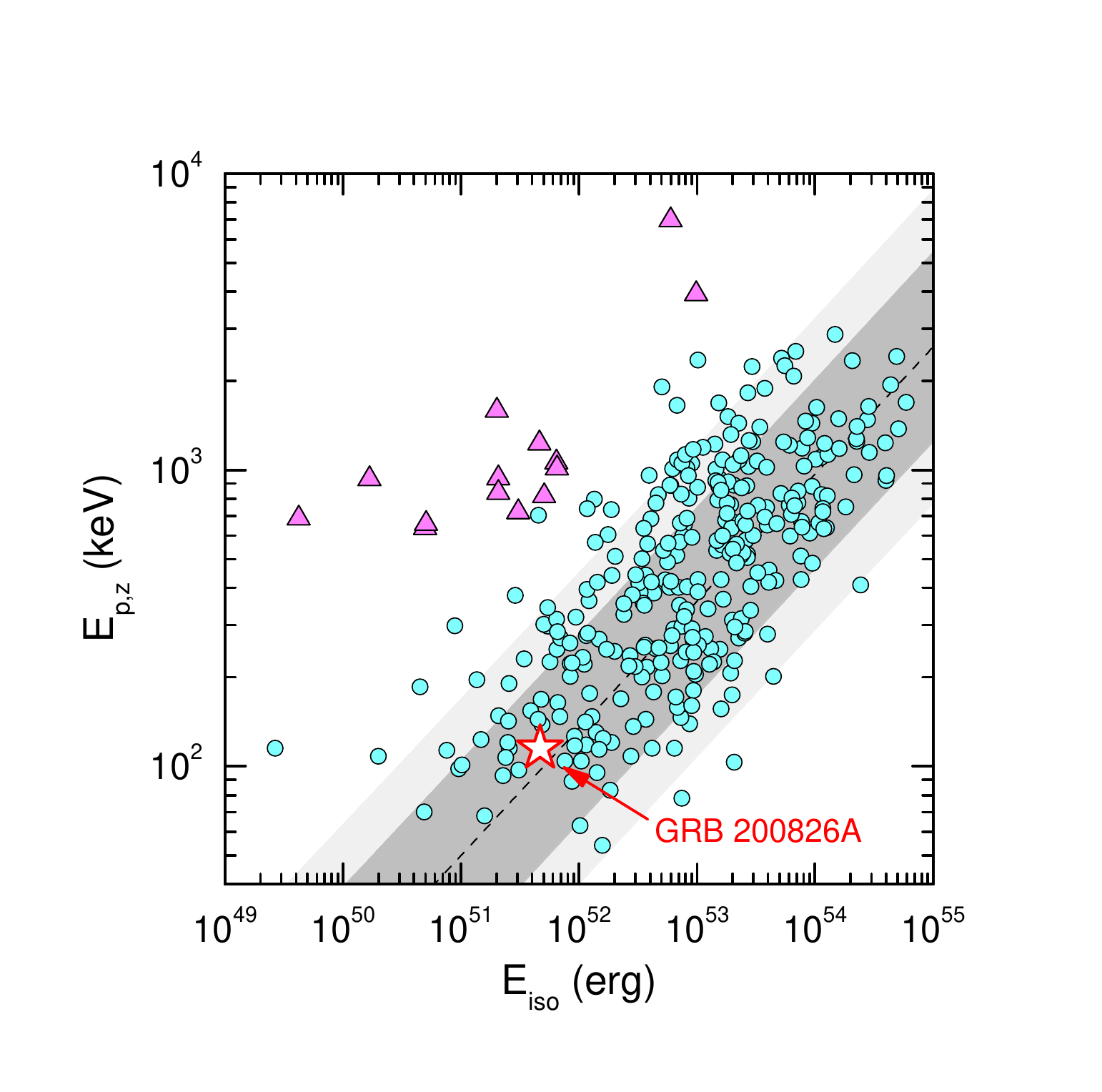}
  \end{minipage} 
\caption{\label{fig:grb_lc} \textbf{The AstroSat and Konus-Wind gamma-ray detections.}
(\textit{left}) Upper pannel panel: The de-trended light curve for GRB 200826A obtained from AstroSat CZTI data. We combined data from all four CZTI quadrants and binned it in 0.05~s bins. We fit and subtract a quadratic trend from the background to obtain zero-mean data. The shaded green region and corresponding green symbols denote a conservative GRB time span excluded from background trend estimation. Similarly red points denote outliers that are automatically flagged and rejected from the background estimate.
Lower panel: A cumulative light curve obtained by summing the de-trended data, and normalised such that the median post-GRB value is 1.0. The dashed horizontal lines denote the 5\% and 95\% intensity levels. The corresponding vertical dotted black lines denote $T_{05}$ and $T_{95}$, yielding $T_{90}$ of $0.96^{+0.71}_{-0.17}$ s.
(\textit{right}) 
Rest-frame energetics of 331 Konus-\textit{Wind} GRBs (SGRB: triangles, LGRB: circles) with known redshift in the $E_\mathrm{iso}$–-$E_\mathrm{peak,z}$ plane, with $E_\mathrm{peak,z}$ the rest frame $E_\mathrm{peak}$. The hardness-intensity (`Amati') relation for LGRBs is
plotted with its 68\% and 90\% prediction
intervals (dark and light gray regions, respectively). GRB\,200826A, as a red star, appears not to be consistent with the SGRB population.}
\end{figure*}

\begin{figure*}
    \centering

\begin{minipage}[b]{0.55\linewidth}
    \centering
`    \includegraphics[width=\textwidth]{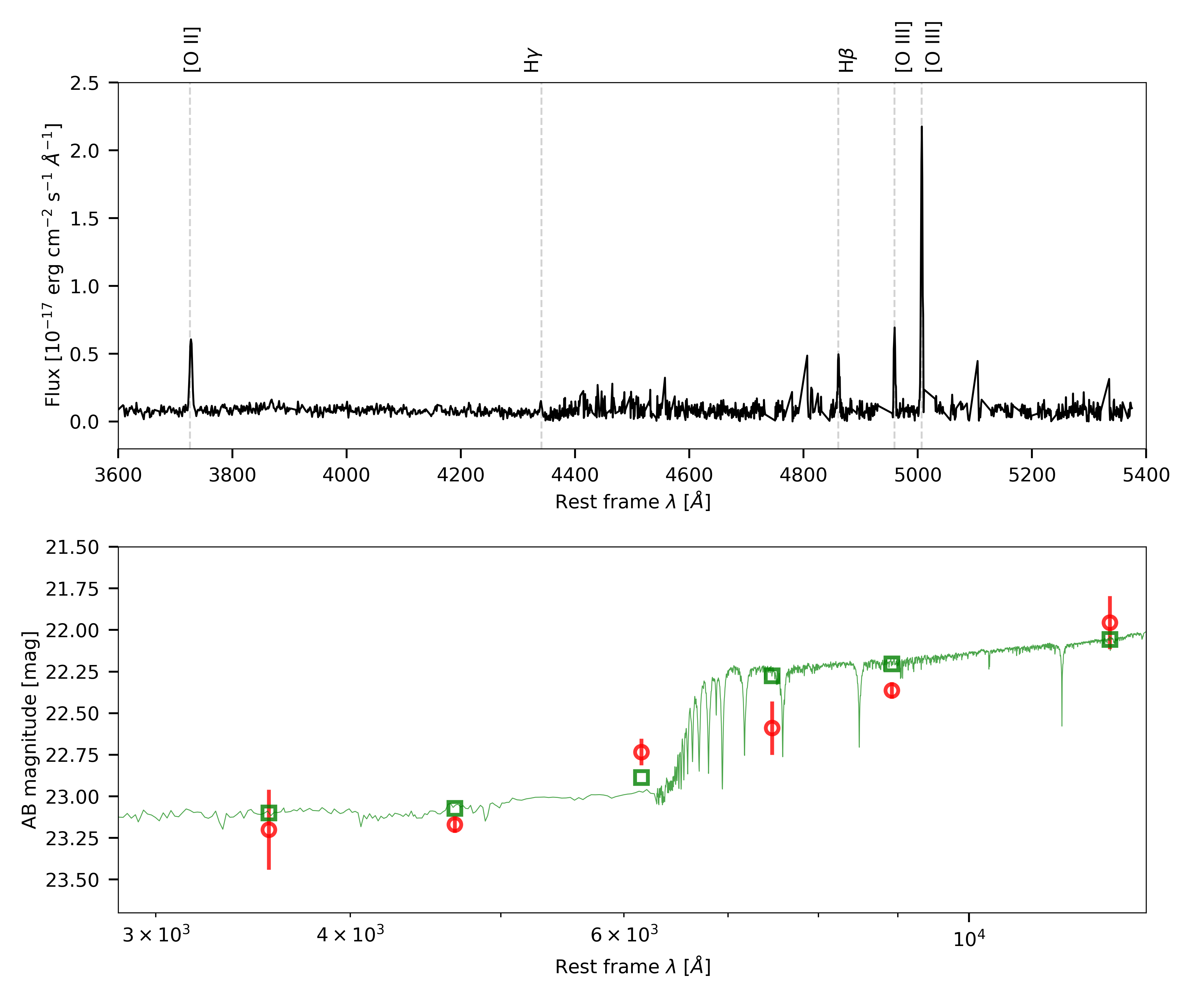}
  \end{minipage}
  \begin{minipage}[b]{0.45\linewidth}
    \centering
\includegraphics[width=\textwidth]{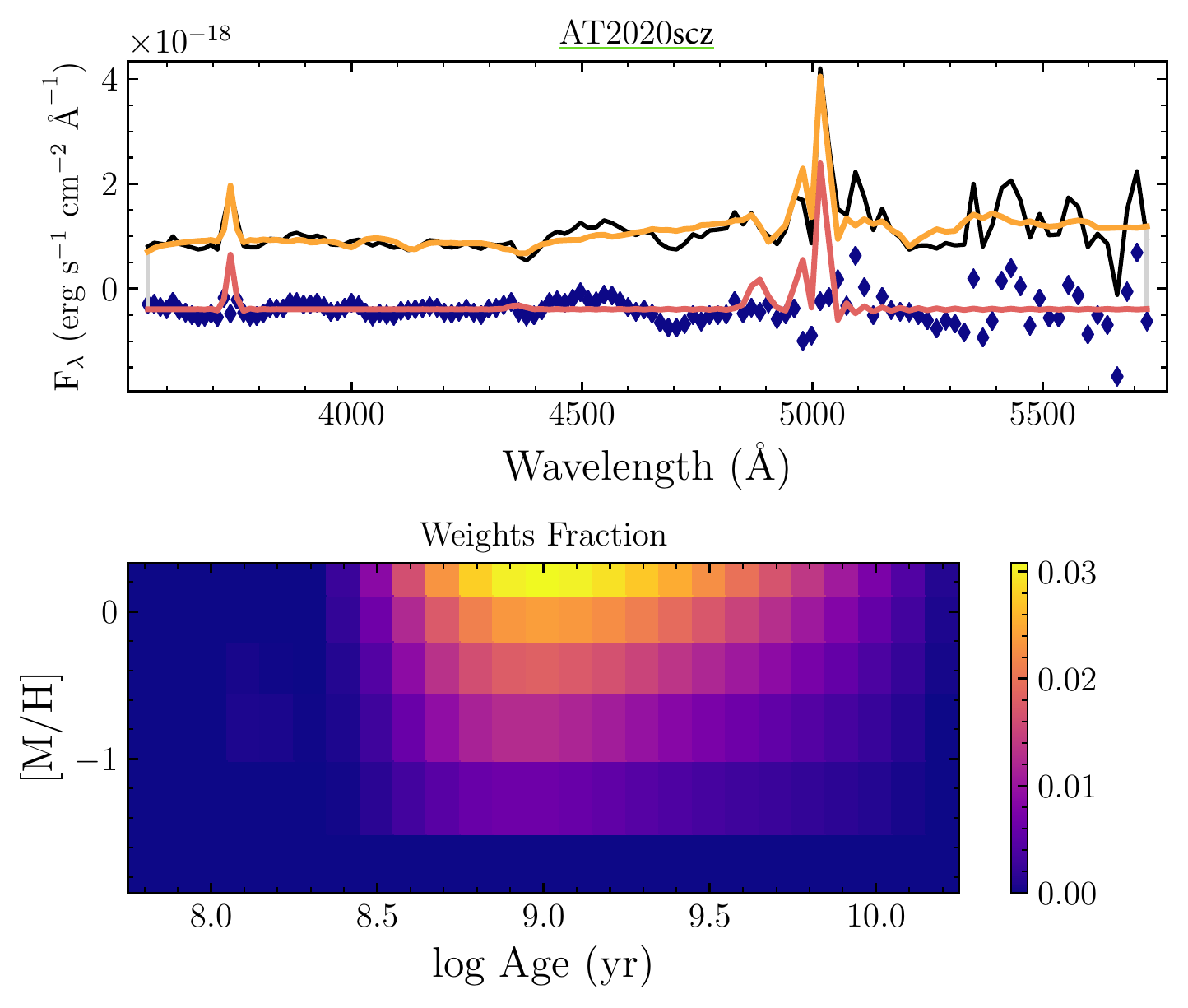}
  \end{minipage} 
  
    \caption{ \textbf{The Host Galaxy.}
    (\textit{left}) In the upper panel, we show the GTC spectrum of the host galaxy and the lines used to determine a redshift of 0.748.
    In the bottom panel, the photometry of the host galaxy (\emph{ugrizJ}, see \ref{table:observations_host}) in the AB system is presented in red circles. The \ac{SED} model and photometry from \texttt{Prospector} are shown in green.
    (\textit{right}) The pPXF host galaxy model results described in §3. (top) The integrated spectrum (black) overlaid with the best-fit spectrum (orange), which sums the contributions of stars and gas in the modeled galaxy. The red spectrum shows the gas contribution to the spectrum, and the blue diamonds show the residuals to the fit. The gas is offset by 1.59e-18 erg s$^{-1}$ cm$^{-2}$. (bottom) The pPXF weights (color bar) of the different stellar population templates used to construct the best-fit galaxy.  }
    \label{fig:galaxySED}
\end{figure*}


\begin{figure}
\includegraphics[width=0.8\textwidth]{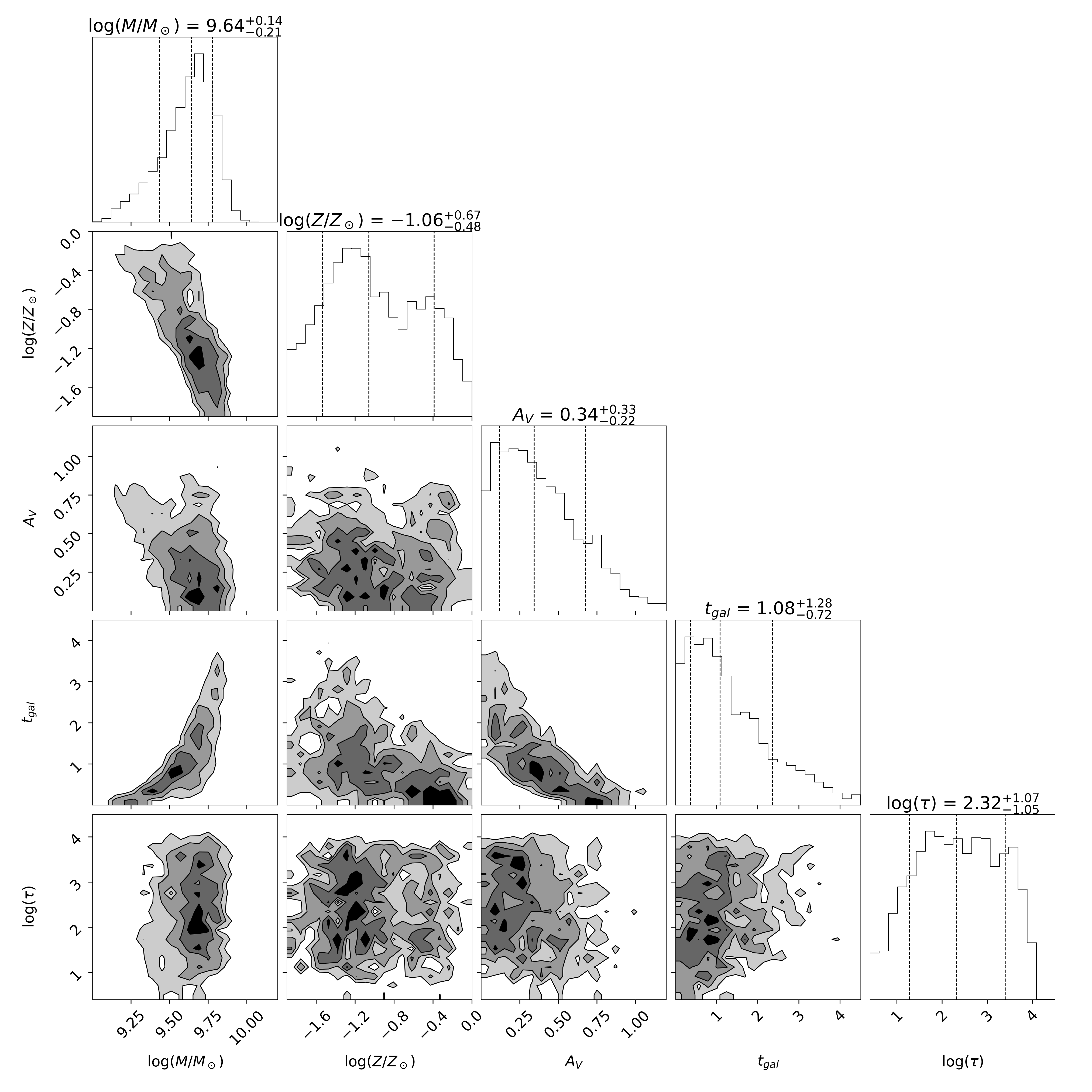}
\caption{ \textbf{Posterior distribution of the \texttt{Prospector} parameters.}
The covariances and posterior probability distributions of the parameters for the host galaxy  derived form the \texttt{Prospector} modelling described in \S \ref{sec:host}.  }
\label{fig:corner_sed}
\end{figure}

\begin{figure*}[!htb]
    \includegraphics[width=\textwidth]{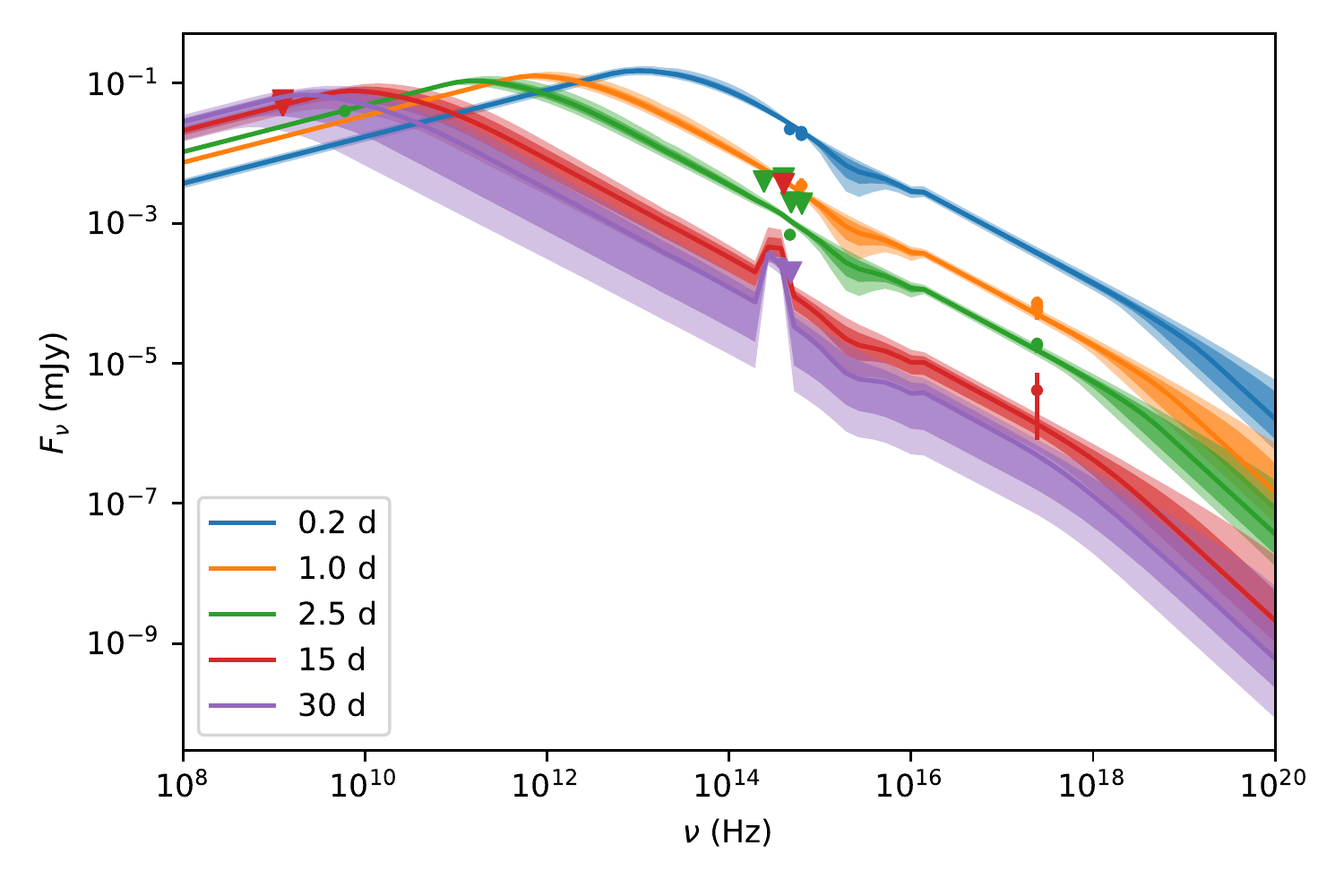}
\caption{    \textbf{\ac{SED} sequence of the afterglow.}
The \ac{SED} of our model compared to observations at five epochs. The cooling frequency $\nu_c$ is located at frequencies higher than 1 keV (i.e. $\nu_c > 2.4 \times 10^{17}$ Hz).
The glitches at optical $\nu \sim 10^{16}$ Hz are the edge of validity of our dust extinction model.  The \ac{SN} makes a large contribution at late times. See the observations in \ref{table:observations_afterglow},\ref{table:observations_xrt}, and \ref{table:observations_radio}.
}
\label{fig:SED_afterglow}
\end{figure*}

\begin{figure}
\includegraphics[width=0.95\textwidth]{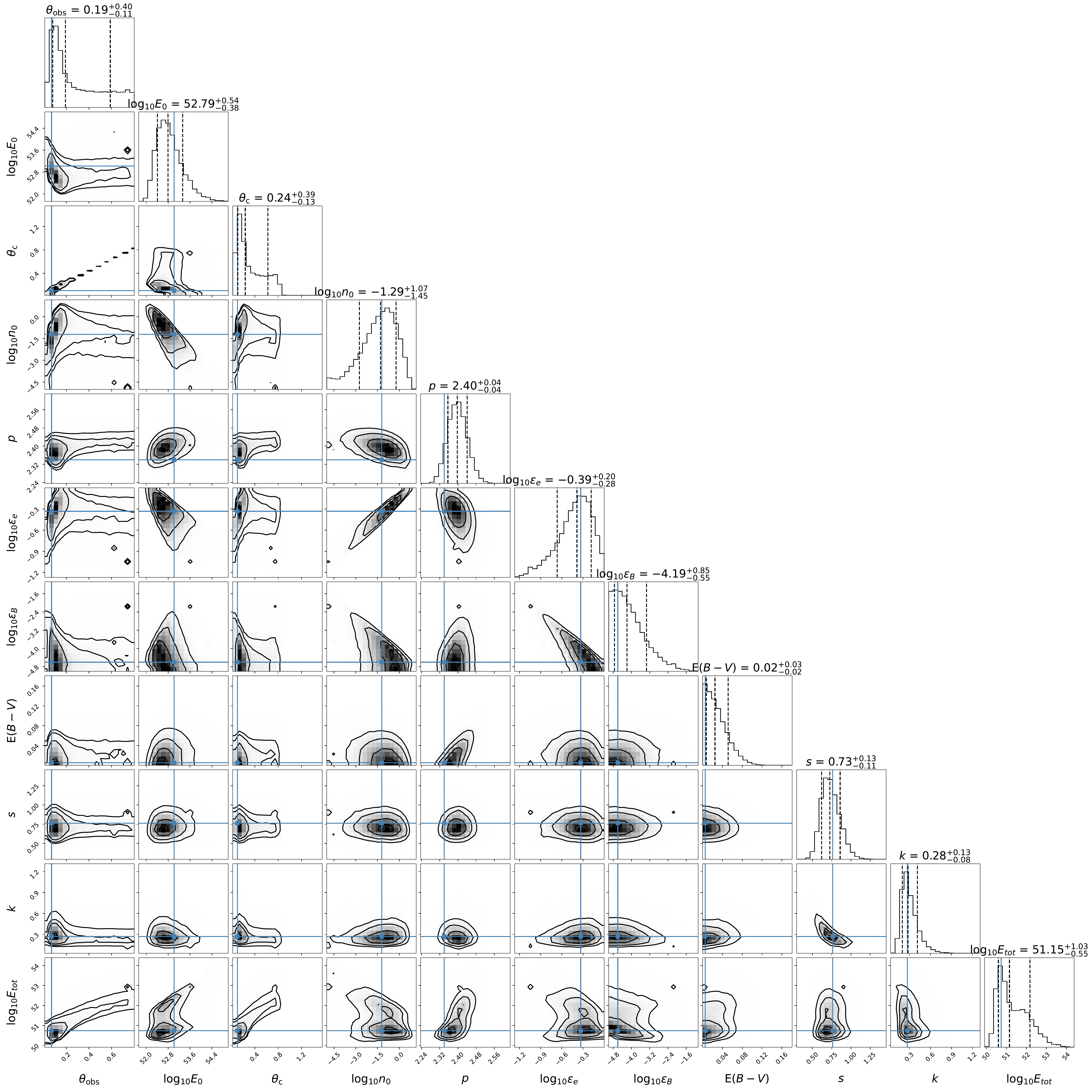}
\caption{\textbf{Posterior distribution for the \texttt{afterglowpy} fit.}
The covariances and posterior probability distributions of the parameters for the top hat jet afterglow with 1998bw-like \ac{SN} model described in \S \ref{sec:afterglow}. The afterglow parameters are the viewing angle $\theta_{\mathrm{obs}}$ (rad), on-axis isotropic kinetic energy $E_0$ (erg), opening angle $\theta_c$ (rad), circumburst number density $n_0$ (cm$^{-3}$), electron spectral index $p$, fraction of energy in accelerated electrons $\epsilon_e$, and fraction of energy in magnetic field $\epsilon_B$. We assume an SMC-like extinction curve with variable $E(B-V)$. The SN model is a 1998bw template with variable stretch $s$ and scale $k$.  The total beaming-corrected kinetic energy in the jet $E_{\mathrm{tot}}$ (erg), computed from the afterglow parameters, is also reported.  The histograms denote the 14, 50, and 84 percentiles of the distributions, with blue lines marking the solution with maximum posterior probability density.}
\label{fig:corner_eps}
\end{figure}

\begin{figure}
    \includegraphics[width=0.8\textwidth]{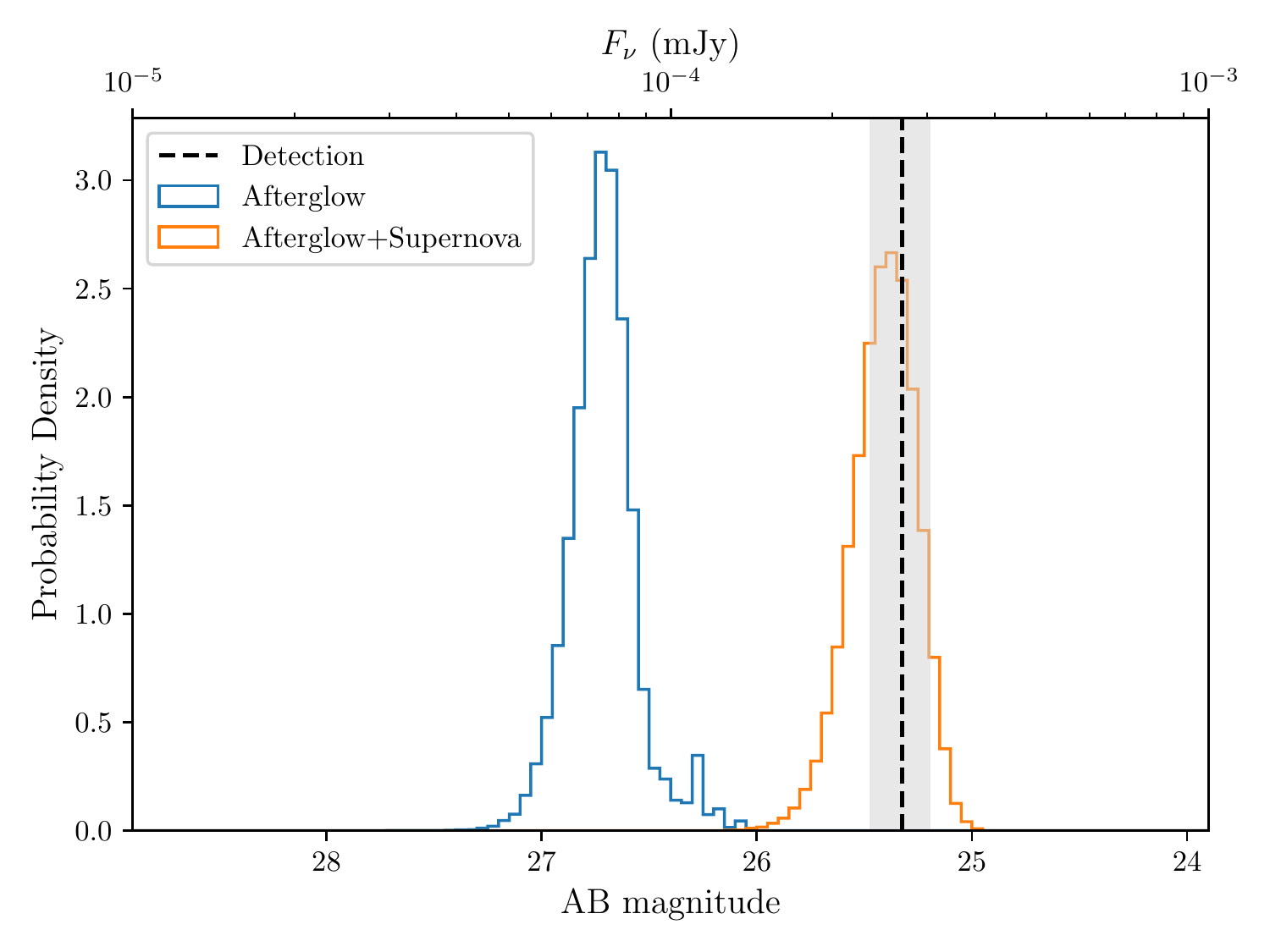}
    \caption{\textbf{Posterior predictive plot for the GMOS $i$-band detection.} The posterior of the AB magnitude estimated at the time of the $i$-band data point ($\sim 28$ days after the trigger) using afterglow only (blue) and afterglow-and-\ac{SN} (orange) light curves are shown.}
    \label{fig:posterior_prediction_afterglow_supernova}
\end{figure}

\newpage
 \begin{table*}
 \resizebox{\textwidth}{!}{
\centering
     \begin{tabular}{cccccccc}
        \hline
        \textbf{Julian day} &\textbf{$\delta t$}&\textbf{Instrument}& \textbf{Filter}&\textbf{AB magnitude} & \textbf{$\sigma_{mag}$} & \textbf{ 5$\sigma$ Limiting magnitude} & \textbf{A$_\nu$}\\
        \hline
         2459085.8814 & -1.81 & P48+ZTF & g & --    &   -- & 21.50 & 0.21\\
         2459085.9678 & -1.72 & P48+ZTF & r & --    &   -- & 21.46 & 0.15\\
         2459086.9011 & -0.79 & P48+ZTF & r & --    &   -- & 21.33 & 0.15\\
         2459086.9696 & -0.72 & P48+ZTF & g & --    &   -- & 21.31 & 0.21\\
         2459087.8986 & 0.21  & P48+ZTF & g & 20.86 & 0.04 & 22.50 &0.21\\
         2459087.9181 & 0.23  & P48+ZTF & r & 20.69 & 0.05 & 22.27 & 0.15\\
         2459087.9651 & 0.28  & P48+ZTF & g & 20.95  & 0.16 & 21.23 & 0.21\\
         2459088.8340 & 1.15  & P48+ZTF & g & 22.75 & 0.26 & 22.53 & 0.21\\
         2459088.9038 & 1.22  & P48+ZTF & r & --    &   -- & 21.30 & 0.15\\
         2459088.9761 & 1.29  & P48+ZTF & g & --    &   -- &  21.2 & 0.21\\
         2459089.6256 & 1.93  & P200+WIRC & J & --  &   -- & 21.8 & 0.06\\
         2459089.9585  & 2.27  & FNT+LCO & r & --    &   -- & 23.30  & 0.15\\
         2459089.9698  & 2.28  & FNT+LCO & g & --    &   -- & 23.41& 0.21\\
          2459090.9200  & 3.23  & LDT+LMI & r & 24.46    &   0.12 & 26.37& 0.15\\
         2459115.9675 & 28.28  & Gemini+GMOS & i &  25.45 & 0.15  & 25.9 &0.11\\
         2459115.9675 & 28.28  & Gemini+GMOS & r &  -- & --  & 25.4 &0.15\\
         2459133.8039 & 46.11  & Gemini+GMOS & i &  -- & --  & 25.5 &0.11\\
         2459133.8039 & 46.11  & Gemini+GMOS & r &  -- & --  & 25.7 &0.15\\
         
         \hline
    \end{tabular}
    }
    \caption{\textbf{Afterglow panchromatic observations.} Observations of the GRB\,200826A afterglow and \ac{SN}. GRB\,200826A was triggered at Julian day 2459087.6874.}
    \label{table:observations_afterglow}
\end{table*}

\newpage
\begin{table*} 
\centering
    \begin{tabular}{ccccccc}
        \hline
        \textbf{Julian day} &\textbf{$\delta t$}&\textbf{Instrument}& \textbf{Filter}&\textbf{Host AB magnitude} & \textbf{ $\sigma_{mag}$}  & \textbf{A$_\nu$}\\ 
        \hline
         2459089.62569 & 9.73  & P200+WIRC & J & 21.11* & 0.16 & 0.05\\
         2459105.80457 & 18.11  & LMI+LDT & u & 23.45   &   0.24  & 0.28 \\
          2459105.80457 & 18.11  & LMI+LDT & g & 23.36    &   0.05  & 0.22\\
          2459105.80457 & 18.11  & LMI+LDT & r & 22.86    &   0.18   & 0.15\\
          2459105.80457 & 18.11  & LMI+LDT & i & 22.66    &   0.16  & 0.11\\
          2459105.80457 & 18.11  & LMI+LDT & z & 22.13    &   0.05  &0.09 \\
          \hline
    \end{tabular}
    \caption{\textbf{Host galaxy panchromatic data.} Observations of the host galaxy of GRB\,200826A. *Magnitudes are in the Vega system.}
    \label{table:observations_host}  
\end{table*}

\newpage
\begin{table*}
\centering
     \begin{tabular}{ccccc}
        \hline
        \textbf{Julian day} &\textbf{$\delta t$}&\textbf{Instrument}& \textbf{Energy range}&\textbf{Count rate [s$^{-1}$]}\\
         \hline
        2459088.386616 & 0.70 & XRT & 0.3–10 keV & $ 2.16 _{-0.53}^{0.53}\times 10^{-2}$\\
        2459088.441940 & 0.75 & XRT & 0.3–10 keV & $ 1.91 _{-0.49}^{0.49}\times 10^{-2}$ \\
        2459088.522328 & 0.83 & XRT & 0.3–10 keV & $ 1.66 _{-0.41}^{0.41}\times 10^{-2}$\\
        2459089.437860 & 1.75 & XRT & 0.3–10 keV & $ 5.64 _{-1.21}^{1.21}\times 10^{-3}$ \\
        2459090.285972 & 2.60 & XRT & 0.3–10 keV & $ 5.36 _{-1.22}^{1.22}\times 10^{-3}$\\
        2459094.090850 & 6.40 & XRT & 0.3–10 keV & $ 3.33 _{-0.77}^{0.77}\times 10^{-3}$ \\
        2459104.352156 & 16.66 & XRT & 0.3–10 keV & $ 1.22 _{-0.67}^{0.98}\times 10^{-3}$\\
        \hline
    \end{tabular}
    \caption{\textbf{Afterglow X-ray detections.} X-ray observations of GRB\,200826A.}
    \label{table:observations_xrt}    
\end{table*}

\newpage
\begin{table*}
\centering
    \begin{tabular}{ccccc}
        \hline
        \textbf{Julian day} &\textbf{$\delta t$}&\textbf{Instrument}& \textbf{Frequency} & \textbf{Flux [erg cm$^{-2}$ s$^{-1}$]}\\
        \hline
        2459089.967407 & 2.28 & VLA & 6 GHz & $ 40 $ **  \\
        2459102.153825 & 14.46 & GMRT & 1.256 GHz &  $<$ 48.6 \\
        2459107.485017 & 19.79 & GMRT & 1.256 GHz &  $<$ 57.4 \\
        \hline
    \end{tabular}
    \caption{\textbf{Radio data.} Radio observations of GRB\,200826A. ** \ac{VLA} data from Ref.\cite{GCN28302}.  }
        \label{table:observations_radio}    
\end{table*}

\newpage
\begin{table*}
\resizebox{\textwidth}{!}{
    \begin{tabular}{ccccccccccc}
        \hline
        \textbf{Time Bins (s)} &\textbf{Model}&\textbf{Amplitude}& \textbf{$E_{peak}$ }&\textbf{$\alpha$} & \textbf{$\beta$} 
        & \textbf{Photon Flux} 
        & \textbf{Photon Fluence} 
        & \textbf{Energy Flux} 
        & \textbf{Energy Fluence} 
        & \textbf{Fit Merit} \\
         & & & \textbf{$[$keV$]$}&  &  
        & \textbf{$[$ph s$^{-1}$ cm$^{-2}]$} 
        & \textbf{$[$ph cm$^{-2}]$ } 
        & \textbf{$[$erg s$^{-1}$ cm$^{-2}]$} 
        & \textbf{$[$erg cm$^{-2}]$} 
        & \\
         \hline
        0.000-0.180 & Comp & 0.3$\pm$0.04 & 114.8$\pm$6.0 & -0.6$\pm$0.1 & - & 32.2$\pm$0.9 & 5.8$\pm$0.2 & 3.1$\pm$0.2$\times 10^{-6}$& 0.6$\pm$0.02 $\times 10^{-6}$& 1.1 \\ 
        0.180-0.318 & Band & 4.2 $\pm$ 2.3 & 61.3$\pm$0.5 & 0.5$\pm$0.3 & -2.7$\pm$0.1 & 41.5$\pm$1.1 & 11.5$\pm$0.2 & 3.7$\pm$0.2$\times 10^{-6}$& 1.2$\pm$0.2$\times 10^{-6}$& 0.9 \\ 
        0.318-0.414 & Band & 0.6 $\pm$ 0.1 & 141.9$\pm$10.3 & -0.5$\pm$0.1 & -3.0$\pm$0.4 & 61.5$\pm$1.5 & 17.4$\pm$0.2 & 7.8$\pm$0.4$\times 10^{-6}$& 1.9$\pm$0.3$\times 10^{-6}$& 0.9 \\ 
        0.414-0.506 & Band & 0.7 $\pm$ 0.1 & 145.4$\pm$13.5 & -0.3$\pm$0.1 & -2.3$\pm$0.2 & 61.4$\pm$1.5 & 23.0$\pm$0.1 & 9.6$\pm$0.4$\times 10^{-6}$& 2.8$\pm$0.4$\times 10^{-6}$& 1.0 \\ 
        0.506-0.607 & Band & 1.2$\pm$0.4 &    99.6$\pm$7.4 & -0.04$\pm$0.2 & -2.5$\pm$0.2 & 56.4$\pm$1.4 & 28.8$\pm$0.1 & 6.9$\pm$0.3$\times 10^{-6}$& 3.5$\pm$0.3$\times 10^{-6}$& 0.9 \\ 
        0.607-0.747 & Band & 0.9$\pm$0.2 &   95.3$\pm$5.3 & -0.2$\pm$0.1 & -3.3$\pm$0.4 & 41.7$\pm$1.1 & 34.6$\pm$0.2 & 4.2$\pm$0.2$\times 10^{-6}$& 4.1$\pm$0.3$\times 10^{-6}$& 0.9 \\ 
        0.747-1.152 & Band & 0.6$\pm$0.5 &   41.2$\pm$3.9 & -0.3$\pm$0.4 & -2.7$\pm$0.2 & 15.0$\pm$0.5 & 40.7$\pm$0.2 & 0.9$\pm$0.1$\times 10^{-6}$& 4.5$\pm$0.3$\times 10^{-6}$& 1.1 \\ 
        \hline
    \end{tabular}
    }
    \caption{\textbf{Parameters of the \emph{Fermi}-\ac{GBM} fit.} Fitting parameter of the \emph{Fermi}-\ac{GBM} gamma-ray spectrum of GRB\,200826A. }
    \label{tab:timeres_gbm}
\end{table*}

\newpage
\begin{table*}
\centering
    \begin{tabular}{cc}
        \hline
        \textbf{Line}  &\textbf{$F_\nu$} \\
         $\AA$ & 10$^{-17}$\,erg\,s$^{-1}$\,cm$^{-2}$\\
        \hline
$[$OII$]_{3726}$ &   1.268 $\pm$   0.37 \\
$[$OII$]_{3729}$ &   2.522 $\pm$  0.37 \\
$[$OIII$]_{5007}$ &  8.21 $\pm$ 3.5 \\
        \hline
        \end{tabular}
    \caption{ \textbf{Host galaxy emission line fluxes.} Fluxes derived with \ac{pPXF} for the lines detected in the \ac{GTC} spectrum of the host galaxy.}
    \label{tab:fluxes}
\end{table*}

\begin{table*}
\centering
    \begin{tabular}{ccc}
        \hline
        \textbf{Parameter} 
        &\textbf{Unit}
        &\textbf{Value}\\
         \hline
        $\theta_v$ & [rad] & $0.20^{+0.53}_{-0.15}$ \\
        $E_{\text{K,iso}}$ & [erg] & $6.0^{+51.3}_{-4.4} \times 10^{52}$ \\
        $\theta_c$ & [rad] & $0.24^{+0.53}_{-0.17}$ \\ 
        $n$ & [cm$^{-3}$] & $5.5^{+187.3}_{-5.4} \times 10^{-2}$ \\
        $p$ &  & $2.40^{+0.07}_{-0.07}$ \\
        $\epsilon_e$ &  & $0.42^{+0.40}_{-0.28}$ \\ 
        $\epsilon_B$ &  & $6.4^{+197.4}_{-5.2} \times 10^{-5}$ \\
        $E(B-V)$    &   & $2.5^{+4.8}_{-2.3} \times 10^{-2}$ \\
        stretch $s$    &   & $0.73^{+0.23}_{-0.16}$  \\
        scale $k$    &   & $0.28^{+0.25}_{-0.12}$ \\
        $E_{\text{k}}$ & [erg] & $1.4^{+54.8}_{-1.2} \times 10^{51}$ \\
        \hline
        \end{tabular}
    \caption{\textbf{Afterglow properties.} Posterior \texttt{afterglowpy} fit model parameters with an SMC extinction curve, a SN1998bw template and including the final Gemini+GMOS detection.  Uncertainties are quoted at 90\%. $E_{\text{k}}$ is the beamed corrected kinetic energy. See $\S$\ref{sec:afterglow} for more details.}
    \label{tab:pars}
\end{table*}

\section{Data Availability}
Upon request, the corresponding author will provide data required to reproduce the figures, including light curves and spectra for any objects. 

\section{Code Availability}

Upon request, the corresponding author will provide code (primarily in python) used to produce the figures.

\bibliographystyleNew{naturemag}
\bibliographyNew{references}

\end{document}